%% file: iclr2027_conference.tex
\documentclass{article} 
\usepackage{iclr2027_conference,times}

\input{math_commands.tex}

\usepackage{hyperref}
\usepackage[utf8]{inputenc} 
\usepackage[T1]{fontenc}    
\usepackage{hyperref}       
\usepackage{url}            
\usepackage{booktabs}       
\usepackage{amsfonts}       
\usepackage{nicefrac}       
\usepackage{microtype}      
\usepackage{xcolor}         
\usepackage[table]{xcolor}   
\usepackage{booktabs}        
\usepackage{amsmath}
\usepackage{enumitem}
\usepackage{caption}
\usepackage{subcaption}
\usepackage{graphicx}
\usepackage[skip=5pt plus 1pt]{parskip}

\title{UniVerse: Benchmarking and Enhancing LALMs on Culturally Inclusive Low-resource Music Understanding}

\author{%
  \normalfont
  Ziya Zhou$^{1,}$\thanks{Equal contribution. $^\dagger$Corresponding authors.} \quad
  Shangda Wu$^{2,}$\footnotemark[1] \quad
  Shenyang Xu$^{2}$ \quad
  Yutong Zheng$^{2}$ \quad
  Dafang Liang$^{2}$ \\
  Suin Chung$^{3}$ \quad
  Danbinaerin Han$^{4}$ \quad
  Junyan Jiang$^{5}$ \quad
  Yongyi Zang$^{2}$ \quad
  Ruibin Yuan$^{1}$ \\
  Rongxiu Zhong$^{7,8}$ \quad
  Shilei Zhang$^{7,8}$ \quad
  Junlan Feng$^{7}$ \quad
  Jinglei Liu$^{9}$ \\
  Haotian Zhou$^{6}$ \quad
  Zijin Li$^{6}$ \quad
  Dasaem Jeong$^{3}$ \quad
  Wei Xue$^{1,\dagger}$ \quad
  Yike Guo$^{1,\dagger}$ \\[8pt]
  $^{1}$HKUST, Hong Kong SAR, China \quad    
  $^{2}$Independent Researcher\\
  $^{3}$Sogang University, Seoul, Korea \quad
  $^{4}$KAIST, Daejeon, Korea \quad
  $^{5}$NYU Shanghai, China\\
  $^{6}$Central Conservatory of Music, Beijing, China\\
  $^{7}$JIUTIAN Research, China Mobile, Beijing, China\\
  $^{8}$The State Key Laboratory of Multimedia Information Processing, Peking University, Beijing, China\\
  $^{9}$China Mobile (Hong Kong) Innovation Research Institute, Hong Kong SAR, China\\
  \texttt{zzchoup@connect.ust.hk}, \texttt{\{weixue, yikeguo\}@ust.hk}
  }

\iclrfinalcopy 
\begin{document}

\maketitle

\input{sections/0_abstract}
\input{sections/1_Intro}
\input{sections/2_RelatedWork}

\input{sections/3_Methodology}
\input{sections/4_Experiments}

\input{sections/5_Conclusion}

\section*{Acknowledgements}
This work was supported in part by the InnoHK Hong Kong Generative AI R\&D Centre (HKGAI). 
This research was also partially funded by the China National Social Science Foundation (Special Project for Rare and Precious Studies: "Construction of a Digital Cultural Heritage Platform for Musical Instruments of Ethnic Minorities in Southwest China") under Grant No.~22VJXG012.

\bibliography{iclr2027_conference}
\bibliographystyle{iclr2027_conference}

\input{sections/Appendix}
\end{document}

%% file: math_commands.tex
\usepackage{amsmath,amsfonts,bm}

\def\eqref#1{equation~\ref{#1}}

\def\1{\bm{1}}

\DeclareMathAlphabet{\mathsfit}{\encodingdefault}{\sfdefault}{m}{sl}
\SetMathAlphabet{\mathsfit}{bold}{\encodingdefault}{\sfdefault}{bx}{n}



%% file: sections/0_abstract.tex
\begin{abstract}

Recent advances in large audio-language models (LALMs) have significantly improved performance in tasks such as music captioning, genre classification, and sound event detection. However, limited attention has been paid to improving their adaptability across diverse musical traditions, particularly folk music rooted in distinct cultural contexts. Folk-music traditions are typically resource-scarce, unevenly represented across regions, and poorly documented. Even when such samples appear in large-scale pre-training, LALMs often fail to capture their structural and stylistic characteristics, partly due to the absence of dedicated evaluation protocols and training solutions. To address these limitations, we introduce \textsc{UniVerse}, a reproducible solution for low-resource music understanding. Specifically, we propose \textsc{UniVerseBench}, a benchmark of 5,042 Q\&A pairs across more than 38 cultural and linguistic entities, constructed via an expert-guided yet highly automated pipeline. In parallel, we construct a fully automated, model-generated multi-turn dialogue training dataset \textsc{UniVerseSet}. By training LALMs on \textsc{UniVerseSet}, we systematically adapt and investigate representative multimodal imbalance learning strategies across both dense and Mixture-of-Experts (MoE) architectures. Experimental results indicate that fully automated data curation combined with imbalance-aware training yields non-trivial improvements, but models still struggle to capture fine-grained acoustic features, indicating a gap between surface-level alignment and deep musical comprehension. The project page and resources are available \href{https://github.com/SylviaZiyaZhou/UniVerse/tree/main}{here}.

\end{abstract}

%% file: sections/1_Intro.tex
\section{Introduction}

Recent advancements in multimodal tokenization and LLMs have significantly improved the performance of Large Audio-Language Models (LALMs) in music understanding and generation. However, they often exhibit a systematic bias toward high-resource genres (e.g., pop, rock) while overlooking the regional modal systems, microtonal scales, and idiomatic rhythms of folk music traditions. Folk music is inherently low-resource, characterized by heterogeneous formats, sparse documentation, and limited annotations. Even within massive pre-training corpora, LALMs fail to adequately capture its structural and stylistic nuances without targeted supervision. Although benchmarks like CMI-bench~\citep{ma2025cmi} have quantified this cultural bias, they remain limited in scope and automation. Moreover, automated data generation, augmentation, and multimodal imbalance learning remain underexplored.

\begin{figure}[htbp] 
    \centering 
    \includegraphics[width=\textwidth]{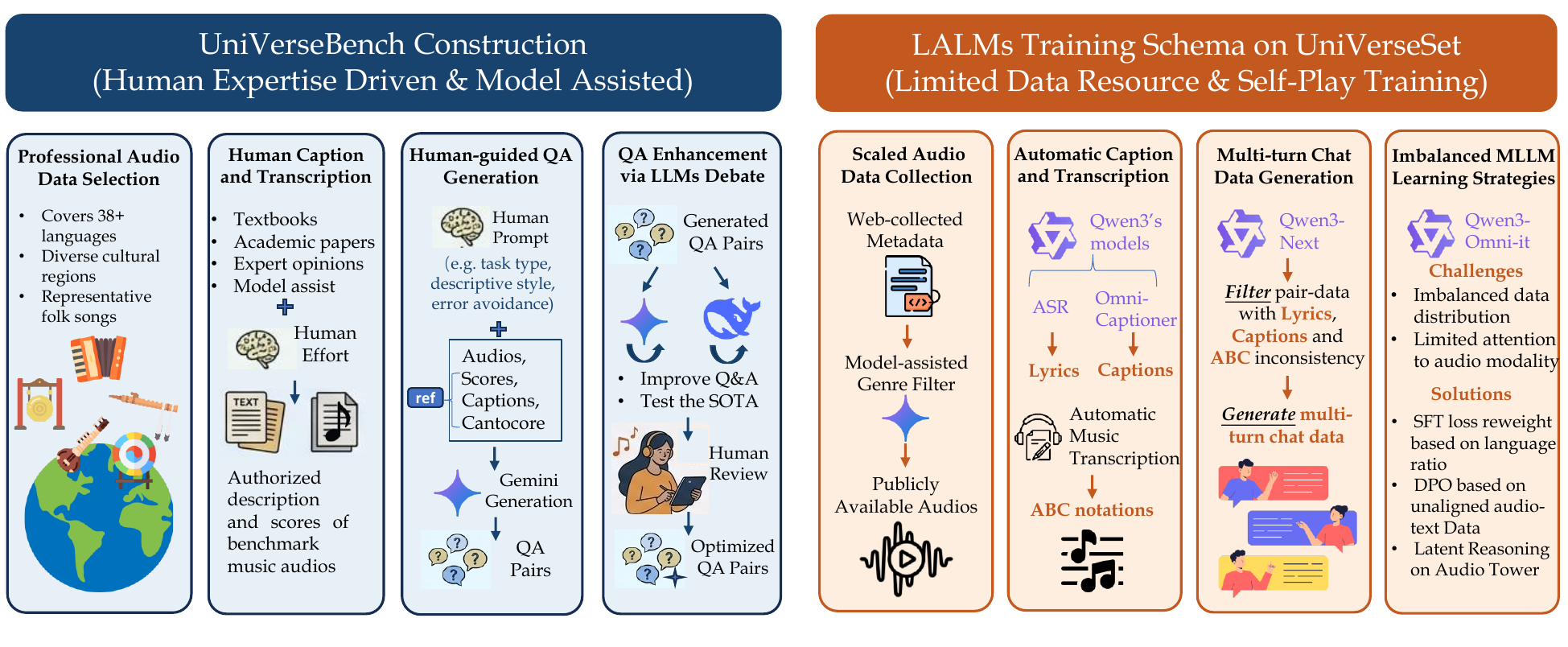} 
    \caption{The pipeline of \textsc{UniVerseBench} construction and LALMs training schema on \textsc{UniVerSet}.} 
    \label{fig:main_review} 
\end{figure}

To explore these issues, we qualitatively probed representative folk music tracks using Gemini 3 Pro~\footnote{https://deepmind.google/models/gemini/pro/}, identifying two kinds of systematic failure: (1) \textbf{Confident misidentification of domain‑specific musical elements with confident hallucination.} LALMs exhibit limited proficiency in identifying both cultural characteristics and low-level acoustic features of regional music. For instance, given an audio recording of a raga, Gemini 3 Pro misclassifies \textit{Raag Abhogi} as \textit{Todi}; it also mistakes the tempo of \textit{Ze wei lao mo} (a piece of \textit{Yi} music from Yunnan Province in China) as a piece of steady 4/4 music, unaware of its mixed rhythms. Such discrepancies may be attributed to three factors that are unconventional symbolic systems (e.g., non-standard tuning), highly variable acoustic environments and various structural patterns like rhythmic variations and pitch bending. These traits are fundamental elements of stylistic identity, and their occurrences poses substantial challenges to model robustness. (2) \textbf{Overreliance on textual context at the expense of acoustic evidence.} Models are overly sensitive to metadata, leading to typical text-driven biases. Most commonly, rich textual contexts induce cross‑cultural interference, misleading the model into mapping acoustic patterns onto incorrect theoretical frameworks (e.g., prompting \textit{"Chinese music"} causes \textit{Raga} audio to be labeled as \textit{Gong Pentatonic} during key identification). Besides, even with correct hints, descriptive metadata can distract the model from objective acoustic analysis (e.g., exposing the location \textit{"Honghe Hani and Yi Autonomous Prefecture"} causes a piece of \textit{Yi} music piece to be misclassified as \textit{Hani} music).

While these vulnerabilities are inherent to LALMs~\citep{kuanCanLargeAudioLanguage2024, sun2025hallucinations, cheng2026aha}, they are acute in low-resource music traditions where data scarcity and text-over-audio reliance amplify models' fragility. To bridge these gaps, we propose \textsc{UniVerse}, including three solutions tailored for cross-cultural, low-resource music understanding:

\begin{itemize}[leftmargin=1.0em]
\item \textbf{A Comprehensive Benchmark of Heterogeneous Contexts:} We introduce \textsc{UniVerseBench}, a benchmark comprising 5,042 QA pairs derived from 372 audio recordings across over 38 languages, with music scores, CantoCore~\citep{savage2012cantocore} features, and aligned captions. The dataset balances expert-curated instances with automated retrievals via a web‑accessible Gemini 3 Pro.

\item \textbf{Human-AI Collaborative Production for Cultural and Musical Diversity:} We propose a reproducible, human-AI collaborative pipeline that integrates diverse cultural elements to balance professional rigor with task viability while preserving authenticity. We further analyze the boundaries of LLM-driven knowledge integration and detail our human-in-the-loop validation.

\item \textbf{Investigation of Post-training Schemas for Low-Resource Audio Understanding:} We systematically investigate training schemas adapted for cross-cultural contexts on \textsc{UniVerseSet}, a training dataset curated via a heterogeneous process distinct from benchmark construction, with lyrics, captions, and scores extracted via open-source LALMs and transcription models. Utilizing these aligned pairs, we evaluate three multimodal imbalance learning strategies: language-aware loss reweighting, direct preference optimization (DPO) on both text and audio towers, and latent reasoning approaches specially designed for dense and MoE architecture.
\end{itemize}

%% file: sections/2_RelatedWork.tex
\section{Related Work}
\subsection{Cross-cultural Music Understanding}

\textbf{Datasets and Benchmarks.} Previous initiatives, such as Compmusic~\citep{zhangCCOMHuQinAnnotatedMultimodal2023, zhouCCMusicOpenDiverse2025, krishnanSanidhaStudioQuality2024, papaioannouDatasetGreekTraditional2022, hanSixDragonsFly2024a}, have established foundations of data resources and benchmarks for cross-cultural analysis. These resources inform instruction-based benchmarks like Audio-Flan~\citep{xue2025audioflan}, MusicQA~\citep{christodoulouMusiQAlDatasetMusic2025}, GlobalMood~\citep{leeGlobalMoodCrossculturalBenchmark2025}, TAU~\citep{linTAUBenchmarkCultural2025} and ArtistMus~\citep{kwonArtistMusGloballyDiverse2025} while specialized studies address attributes such as emotion, timbre, and rhythm. However, these benchmarks often rely on culturally bounded taxonomies and lack regional diversity. Their unique construction processes frequently hinder reusability in other low-resource domains, limiting their utility for evaluating model generalization to underrepresented folk traditions.


\textbf{Training and Evaluation.} Recent efforts seek to enhance auditory understanding via continual pre-training on multicultural data~\citep{kanatasCultureMERTContinualPreTraining2025}, multimodal alignment through contrastive or generative objectives~\citep{wuCLaMP2Multimodal2025}, and supervised fine-tuning (SFT) for descriptive tasks~\citep{papaioannouUniversalMusicRepresentations2025}. Nevertheless, evaluation remains focused on standard semantic attributes (e.g., genre, mood) rather than nuanced features like microtonal scales or ritualistic contexts. Consequently, the capacity of audio-language models to capture these neglected traits remains underexplored.

\subsection{Post-training of MLLMs}

\textbf{Multimodal Perception beyond Text.} Vision-language frameworks routinely decouple perception from reasoning to enhance cultural sensitivity~\citep{qiaoPrismFrameworkDecoupling, romeroCVQACulturallydiverseMultilingual2024, liuCultureVLMCharacterizingImproving2025}. In contrast, Large Audio-Language Models (LALMs) still suffer from acoustic hallucinations regarding object existence and temporal order~\citep{kuanCanLargeAudioLanguage2024}, while the scarcity of text descriptions in low-resource traditions hinders purely text-supervised generalization. Because existing music QA benchmarks predominantly evaluate text reasoning over auditory perception~\citep{zangAreYouReally2025}, \textsc{UniVerse} minimizes prompt textual complexity to directly target intrinsic audio variations.

\textbf{Imbalanced Learning in Low-Resource Scenarios.} Cross-cultural music understanding faces long-tailed language distributions and text-dominance biases. To address these bottlenecks, we build upon three post-training paradigms: (1) \emph{Language-weighted loss} to rebalance long-tailed supervision without data resampling~\citep{li2024languageimbalance}; (2) \emph{Text and audio preference optimization (DPO)} to reinforce musical grounding over text priors by contrasting matched versus mismatched audio inputs~\citep{rafailov2023dpo, moddpo2026}; and (3) \emph{Encoder-side latent representation alignment (REPA)}, which utilizes continuous latent states rather than explicit text tokens to capture subtle musical structures across dense and MoE backbones~\citep{dai2026latentomni, xu2025qwen3omni}.


%% file: sections/3_Methodology.tex
\section{Methodology}
Figure~\ref{fig:main_review} illustrates the pipeline for constructing \textsc{UniVerseBench} alongside the training schema applied to \textsc{UniVerseSet}. To prevent data leakage, we strictly isolate the two processes to ensure zero overlap in their data sources. Details are illustrated in Section~\ref{section3.1} and Section~\ref{section3.2}.

\subsection{UniVerseBench Construction: From Local Traditions to Global Scales}\label{section3.1}

\begin{figure*}[t]
  \centering
  \includegraphics[width=\textwidth]{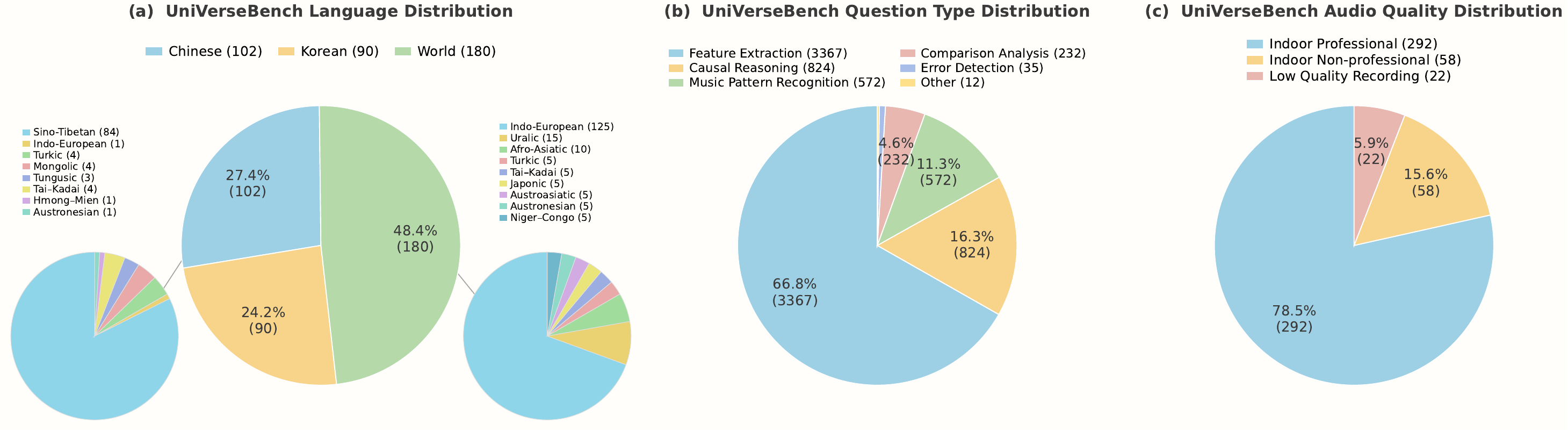}
  \caption{Composition of \textsc{UniVerseBench}.
  \textbf{(a)~Language distribution};
  Chinese and World slices are further broken down by language family
  (Chinese-region labels from metadata language/ISO codes;
  World languages mapped to macro families).
  \textbf{(b)~Question-type distribution} over the final benchmark: feature extraction, causal reasoning,
  music pattern recognition, comparison analysis,
  error detection, other (includes music completion, etc.).
  \textbf{(c)~Audio Source and Quality distribution} from the
  quality audit sample: indoor professional, indoor non-professional,
  and low-quality recording.}
  \label{fig:universe-composition}
\end{figure*}

\subsubsection{Multi-Source Content Curation and Annotation}
Figure~\ref{fig:universe-composition} shows the basic components of \textsc{UniVerseBench}. The \textit{world} section's music inherits from the VoC~\cite{wu2026voicescivilizationsmultilingualqa}, a previous multilingual QA benchmark for global music understanding. To balance scholarly rigor, public relevance, and cultural authenticity, we designed two strategies: expert-led selection and multimodal annotation:

\begin{itemize}[leftmargin=1.0em]
    \item \textbf{Expert-Led Selection}: A panel of ethnomusicology experts curated the initial corpus by referencing authoritative folk music textbooks and integrating trends from global streaming platforms. In order to preserve "living heritage", we incorporated data from regional government-led digital platforms and high-fidelity field recordings from cultural organizations.
    \item \textbf{Multimodal Annotation}: For each selected track, musicology experts provided initial text descriptions based on academic references, which were subsequently augmented by Gemini 3 Pro to enhance linguistic richness. Experts in performance and composition then manually transcribed the music to provide precise musical scores in the form of ABC notation. To extract the features of global vocal traditions, we then apply CantoCore~\citep{savage2012cantocore}, a computational ethnomusicological framework, to the transcribed ABC notation scores.  
\end{itemize}

\begin{figure*}[htbp]
  \centering
  \begin{subfigure}[t]{0.52\textwidth}
    \centering
    \includegraphics[width=\linewidth]{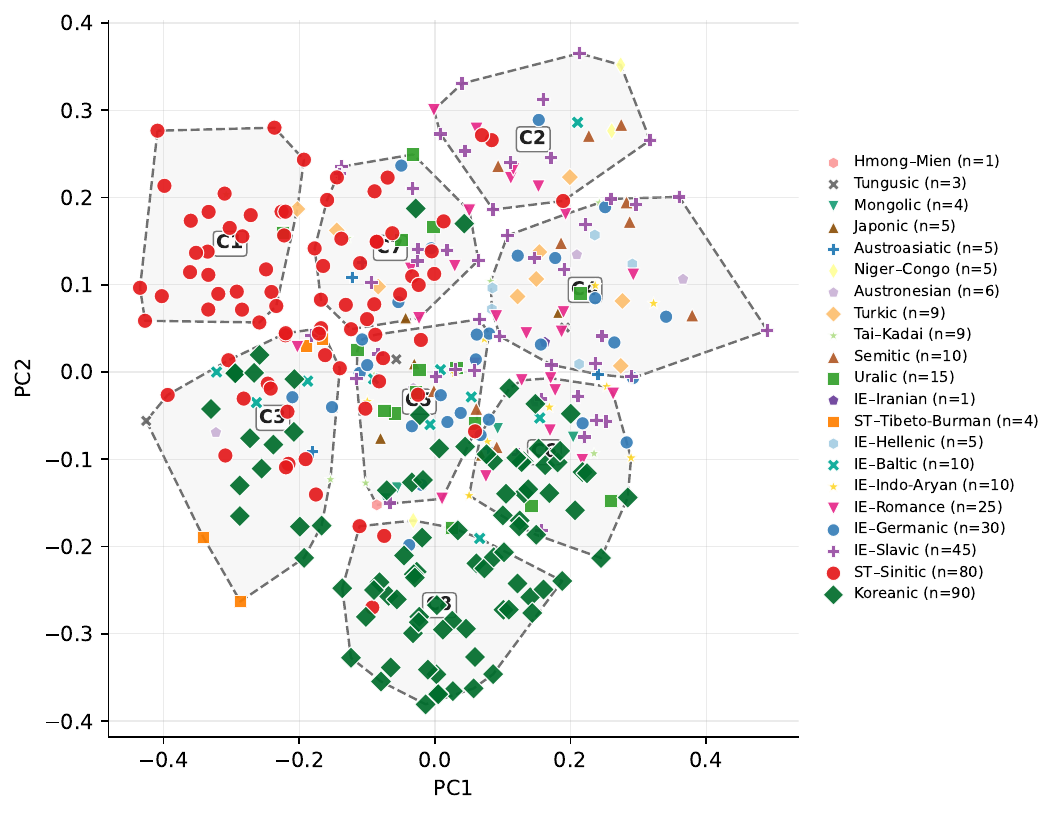}
    \caption{Audio embeddings by language family.}
    \label{fig:audio-lang-pca}
  \end{subfigure}\hfill
  \begin{subfigure}[t]{0.48\textwidth}
    \centering
    \includegraphics[width=\linewidth]{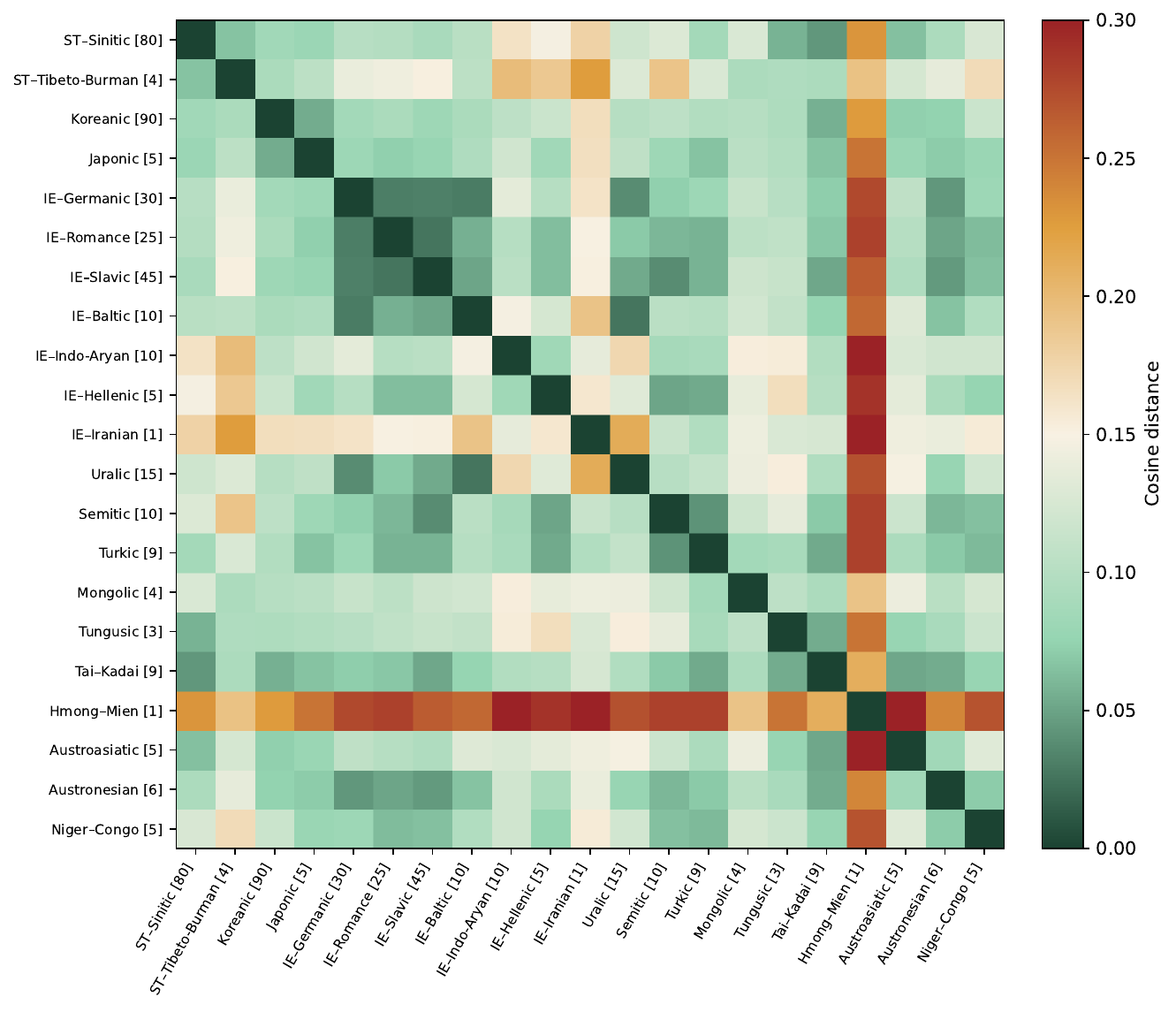}
    \caption{Centroid cosine distance matrix of the centroid cosine distance.}
    \label{fig:audio-lang-centroid}
  \end{subfigure}
  \caption{Acoustic structure of the benchmark audio excerpts in Qwen2.5-Omni embedding space.
  Points in panel (a) are colored by language family and each cluster's convex hull is outlined by dashed polygons;
  Sino-Tibetan branches are labeled \texttt{ST--Sinitic} / \texttt{ST--Tibeto-Burman} (parallel to \texttt{IE--*})).
  Panel (b) shows cosine distances between language family centroids.}
  \label{fig:audio-language-structure}
\end{figure*}

\subsubsection{Instruction Design and QA Generation}
We assessed the reasoning and instruction-following limits of LALMs by constructing a specialized set of evaluation tasks. The structure of each question-answer (QA) pair is defined as follows: given a question $Q$, options $O$, an audio clip $X$ and its associated metadata or prior knowledge $K$, a QA pair $Q_{pair}$ is defined as a tuple:$$Q_{pair} = (Q, O, X, K, A_{ns}, \tau, \lambda, \delta)$$where $A_{ns}$ denotes the answer, $\tau$ denotes the task type (e.g., rhythm analysis, cultural classification); $\delta \in \{0, 1\}$ is a text-dependency label, indicating whether the question requires external textual context ($\delta=1$) or solely relies on acoustic signals ($\delta=0$). This labeling schema allows for the decoupled detection of a model's inherent audio understanding versus its reliance on linguistic cues.

\subsubsection{Generated QA Pairs Optimization and Verification}
To ensure the quality of our proposed benchmark and prevent text-only shortcuts, we implement an automated optimization module followed by a human-AI verification protocol.

\paragraph{Questions and Options Optimization} We first iteratively refine questions for which Gemini's output does not match the music expert's reference, creating semantically complicated items that challenge higher-order musical reasoning~\citep{grari2026agentic}. Then we filter out text-only shortcuts by computing a Perceptual Index (PI) using a text-only LLM~\citep{zangAreYouReally2025}. Questions with low PI—whose answer can be guessed without listening—are flagged, and their distractor options are regenerated to be musically plausible but impossible determine via text alone.


\paragraph{Hybrid Verification} We employ a randomized human-AI consensus protocol to verify ground-truth accuracy. Experts assess each QA pair output through a multi-turn review process informed by the ABC notation, captions, and CantoCore features. A QA pair is accepted when at least two initial experts agree; otherwise, an additional expert independently reviews the item with the assistance from Gemini 3 Pro. The pair is then retained only if this expert's judgement agrees with either Gemini's output or the consensus of initial experts, otherwise it is discarded. This cross-verification procedure minimizes both individual bias and model hallucination. Appendix~\ref{app:universe_benchamrk_verification} provides the automated workflows and prompt templates.


\subsection{Post-training Framework for Low-Resource Music Understanding}\label{section3.2}

\subsubsection{UniVerseSet: Automated Data Construction Pipeline}\label{sec:data_pipeline}
To build a scalable training dataset for traditional music understanding, we design a three-stage pipeline encompassing raw data collection, feature extraction, and multi-turn dialogue generation:

\textbf{Data Collection and Processing.} We gather culturally diverse audio tracks in three sequential steps:
\begin{itemize}[leftmargin=1.0em]
    \item \textit{Metadata Curation:} Retrieving track metadata from a global music streaming service to extract a curated list of track titles and their corresponding artist names.
    \item \textit{Cross-Platform Alignment:} Aligning the extracted title-artist pairs to locate their linked YouTube identifiers.
    \item \textit{Audio Retrieval:} Downloading the raw audio tracks using these identifiers and discarding excessively long tracks longer than 10 minutes.
\end{itemize}

\textbf{Automated Feature Annotation.} Because raw waveforms lack structured semantic or musical details, we extract multi-aspect textual annotations spanning symbolic, linguistic, and acoustic domains:
\begin{itemize}[leftmargin=1.0em]
    \item \textit{Music Scores:} Using an internally trained version of SheetSage~\citep{donahue2022melody} (which will be open-sourced in the future) to transcribe melody contours and chord progressions into ABC notation.
    \item \textit{Lyrics:} Employing Qwen3-ASR~\citep{shi2026qwen3asr} to transcribe lyrics and predict the language to preserve narrative context.
    \item \textit{Captions:} Utilizing Qwen3-Omni-Captioner~\citep{ma2025omni} to generate detailed descriptions of fine-grained acoustic attributes (e.g., timbre, instrumentation).
\end{itemize}
To ensure data quality, we cross-validate the transcribed modalities against the raw metadata using Qwen3-Next-80B-A3B-Instruct\footnote{https://huggingface.co/Qwen/Qwen3-Next-80B-A3B-Instruct}. Tracks are excluded if the LLM identifies significant cross-modal inconsistencies.

\textbf{Dialogue Synthesis.} We generate multi-turn conversations by prompting Qwen3-Next-80B-A3B-Instruct to simulate interactive listening sessions. Firstly, we construct structured user profiles (defining demographic, linguistic, and task parameters) for each track. Both the user and the assistant are framed as blind listeners. Then We apply constraints to prevent answer leakage in user queries and eliminate factual hallucinations by enforcing step-by-step thinking traces in the assistant's reasoning.

Through this automated pipeline, we compile our final instruction-following dataset consisting of 113,023 multi-turn dialogues, totaling 510,078 QA pairs (averaging 4.51 turns per session). The dataset spans 36 unique languages, capturing a multilingual long-tail distribution of global music heritages (Figure~\ref{fig:lang_dist}). Detailed specifications regarding platform configurations, transcription heuristics, user profile definitions, and multi-turn dialogue constraints are deferred to Appendix~\ref{app:data_collection_details}, \ref{app:feature_labeling_details}, and \ref{app:dialogue_synthesis_details}.

\subsubsection{Low-resource Multimodal Learning Strategies}\label{sec:training_strategy}

\begin{figure}[t]
    \centering
    \includegraphics[width=1\linewidth]{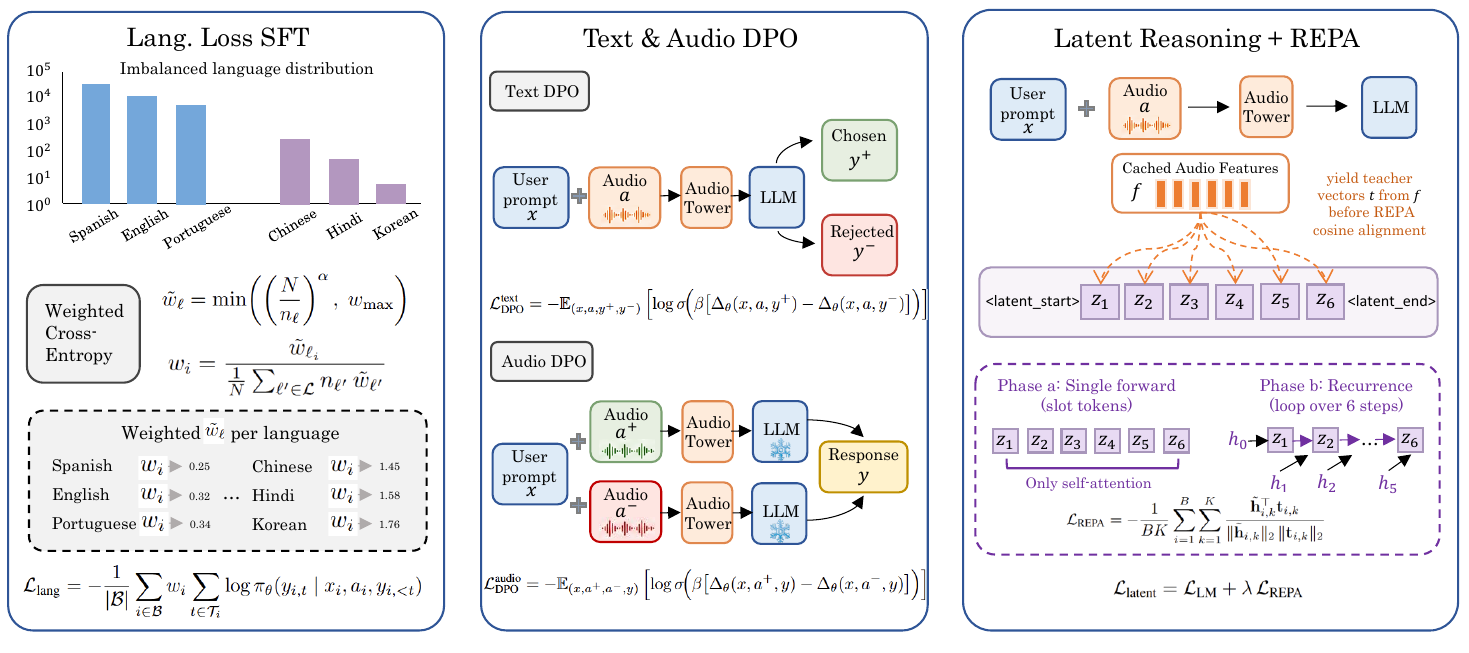}
    \caption{Overview of the three post-training strategies.
    \textbf{Lang.\ Loss SFT} reweights the supervised loss by language frequency to boost low-resource traditions.
    \textbf{Text~/~Audio DPO} applies complementary preference stages post-SFT, optimizing text responses under fixed audio and audio inputs under fixed responses.
    \textbf{Latent Reasoning + REPA} provides architecture-specific alignment: latent reasoning with REPA on dense Qwen2.5-Omni-7B (Qwen2.5-Omni), and encoder-side REPA with a frozen decoder on MoE Qwen3-Omni-30B-A3B-Instruct (Qwen3-Omni).}
    \label{fig:training_strategies}
\end{figure}

\textbf{Language-reweighted SFT} extends standard multimodal SFT~\citep{cocchiLLaVAMOREComparativeStudy2025, zhuInternVL3ExploringAdvanced2025} with per-language loss weights~\citep{li2024languageimbalance, liu2024plug}, up-weighting long-tail languages in training data \emph{without} resampling. 
Training data is highly imbalanced across clip languages. Let each training example $i$ be associated with a language label $\ell_i \in \mathcal{L}$, with corpus counts $n_\ell = \sum_i [\ell_i=\ell]$ and $N=\sum_\ell n_\ell$. To address this, we assign a smooth inverse-frequency weight to each language:
\begin{equation}
    \tilde{w}_\ell = \min\!\left(\left(\frac{N}{n_\ell}\right)^{\alpha},\; w_{\max}\right),
\end{equation}
where $\alpha$ is a smoothing factor and $w_{\max}$ caps extreme weights. We then normalize $\tilde{w}_\ell$ to ensure the expected sample weight equals 1:
\begin{equation}
    w_i = \frac{\tilde{w}_{\ell_i}}{\frac{1}{N}\sum_{\ell' \in \mathcal{L}} n_{\ell'}\,\tilde{w}_{\ell'}}, \qquad \text{such that} \quad \mathbb{E}_i[w_i]=1.
\end{equation}
During supervised fine-tuning on audio-conditioned dialogues $(x_i, a_i, y_i)$, we optimize a language-reweighted cross-entropy loss over the target tokens $\mathcal{T}_i$:
\begin{equation}
    \mathcal{L}_{\text{lang}} = -\frac{1}{|\mathcal{B}|}\sum_{i \in \mathcal{B}} w_i \sum_{t \in \mathcal{T}_i} \log \pi_\theta\!\left(y_{i,t}\mid x_i, a_i, y_{i,<t}\right),
\end{equation}
where $\mathcal{B}$ denotes the mini-batch. This approach scales the gradient contributions of underrepresented languages without data resampling, balancing the effective training distribution while preserving the full multilingual coverage.

\textbf{Text and audio-tower DPO} applies DPO~\citep{rafailov2023dpo} to music-conditioned pairs~\citep{wang2024mdpo, moddpo2026}: text preference is expected to improve culturally grounded answers under the same clip; audio preference, inspired by input-contrastive audio alignment~\citep{moddpo2026, acpo2026}, penalizes answers invariant to swapped music while keeping the LLM frozen. Following Lang Loss SFT, we apply preference optimization from a frozen reference $\pi_{\text{ref}}$.
For a response $y$, define
\begin{equation}
    \Delta_\theta(x, a, y)
    = \log \pi_\theta(y \mid x, a) - \log \pi_{\text{ref}}(y \mid x, a).
\end{equation}
\emph{Text DPO} uses pairs $(x_i,a_i,y_i^{+},y_i^{-})$ with fixed aligned audio $a_i$ and a dispreferred response $y_i^{-}$ (e.g., text-only or incorrect answers), and minimizes
\begin{equation}
    \mathcal{L}_{\text{DPO}}^{\text{text}}
    = -\mathbb{E}
    \log \sigma\!\Big(
    \beta \big[
    \Delta_\theta(x, a, y^{+}) - \Delta_\theta(x, a, y^{-})
    \big]
    \Big).
\end{equation}
The LLM and audio tower are updated jointly under fixed acoustic conditioning.
This stage is used for both dense and MoE backbones.

\emph{Audio-tower DPO} is applied only on the dense backbone.
Following conditional modality-contrast preference learning~\citep{wang2024mdpo,chen2026omnidpo} with an RPO-style chosen-side likelihood anchor~\citep{pang2024rpo}, we keep $y_i$ fixed and contrast aligned audio $a_i^{+}$ against a hard-negative swap $a_i^{-}$ (silence at most sparsely), freeze the LLM, and update only the audio tower with $\alpha>0$:
\begin{equation}
    \mathcal{L}_{\text{DPO}}^{\text{audio}}
    = -\mathbb{E}
    \log \sigma\!\Big(
    \beta \big[
    \Delta_\theta(x, a^{+}, y) - \Delta_\theta(x, a^{-}, y)
    \big]
    \Big)
    + \alpha\,\mathcal{L}_{\text{NLL}}(y \mid x, a^{+}).
\end{equation}

On MoE models, this recipe fails to stabilize. Without RPO, audio-only DPO suffers from loss collapse (near-zero loss and deeply negative rewards across both branches). Adding dense-model stabilization---RPO, reduced learning rates or $\beta$, gradient clipping, or unfreezing router gates---instead triggers early gradient norm explosion (NaNs). We thus restrict MoE acoustic adaptation to supervised cross-entropy ($\mathcal{L}_{\text{CE}}$ and $\Delta\text{CE}$), limiting audio-tower DPO exclusively to dense models.

\textbf{REPA-Grounded Latent Learning.} We extend REPA~\citep{yu2025representationalignmentgenerationtraining} and latent multimodal reasoning~\citep{hao2024coconut, jeon2026vision, dai2026latentomni} to long musical inputs using architecture-specific designs: decoder continuous slots on dense Qwen2.5-Omni and encoder-side $K$-step pooling on MoE Qwen3-Omni~\citep{tang2024salmonn, xu2025qwen3omni}, preserving standard inference with a frozen Thinker. To overcome standard SFT's under-utilization of long acoustic context in extended chain-of-thought, we introduce $K$ continuous latent reasoning steps aligned to offline audio-encoder representations via REPA.
For each example $(x_i,a_i,y_i)$ we minimize
\begin{equation}
    \mathcal{L}_{\text{latent}} = \mathcal{L}_{\text{LM}} + \lambda\,\mathcal{L}_{\text{REPA}},
    \qquad \lambda > 0,
\end{equation}
where $\mathcal{L}_{\text{LM}}$ is next-token cross-entropy on the assistant response $y_i$.
At step $k\in\{1,\ldots,K\}$, REPA compares a trainable student $\mathbf{s}_{i,k}\in\mathbb{R}^{d_s}$ with a frozen teacher $\mathbf{t}_{i,k}\in\mathbb{R}^{d_a}$.
Teachers are obtained offline: frame-level features $\mathbf{f}_i\in\mathbb{R}^{T_i\times d_a}$ are cached for audio $a_i$ and compressed by segment-wise average pooling so $\mathbf{t}_{i,k} = \mathrm{Pool}_K(\mathbf{f}_i)_k.$
Students $\{\mathbf{s}_{i,k}\}$ are computed during training, and they differ between architectures.
When $d_s\neq d_a$, $\mathbf{s}_{i,k}$ is linearly mapped to $\mathbb{R}^{d_a}$ as $\tilde{\mathbf{s}}_{i,k}=\mathbf{W}\mathbf{s}_{i,k}$; otherwise $\tilde{\mathbf{s}}_{i,k}=\mathbf{s}_{i,k}$.
With $\ell_2$ normalization on both sides, we define:
\begin{equation}
    \mathcal{L}_{\text{REPA}}
    = -\frac{1}{BK}\sum_{i=1}^{B}\sum_{k=1}^{K}
    \frac{\tilde{\mathbf{s}}_{i,k}^{\top}\mathbf{t}_{i,k}}
    {\|\tilde{\mathbf{s}}_{i,k}\|_2\,\|\mathbf{t}_{i,k}\|_2},
\end{equation}

where $B$ is the batch size and no gradient flows through $\mathbf{f}_i$.

\medskip
\textbf{Dense Omni Models.} For Qwen2.5-Omni, joint LLM--audio tuning and multi-pass decoding remain feasible. The target $y_i$ is augmented as
\begin{equation}
    \hat{y}_i = \texttt{<latent\_start>}\, z_1 \cdots z_K\, \texttt{<latent\_end>}\, y_i,
\end{equation}
where $z_k$ are label-masked slots with student vectors $\mathbf{s}_{i,k} \equiv \mathbf{h}_{i,k}\in\mathbb{R}^{d_h}$ (final-layer decoder states). Training has two phases: \emph{Phase~a} inserts learnable \texttt{<latent\_slot>} embeddings in one pass; \emph{Phase~b} replaces slots with \texttt{<latent\_rec>} and applies latent recurrence where input embedding at $z_k$ is replaced by preceding state $\mathbf{h}_{z_{k-1}}$ ($z_0\equiv\texttt{<latent\_start>}$), requiring an extra pass. This evolves latents inside the LLM, anchored to audio via REPA.

\medskip
\textbf{MoE Omni Models.} For Qwen3-Omni, recurrent latent decoding causes prohibitive overhead and breaks single-forward serving (e.g., vLLM); slot REPA on routed states was also unstable. We thus shift supervision to the \emph{audio-conditioning path}: freezing the MoE decoder while updating only the audio tower and router gates under $\mathcal{L}_{\text{LM}}+\lambda\mathcal{L}_{\text{REPA}}$. From audio activations $\mathbf{g}_i$, students are $K$-step pooled ($\mathbf{s}_{i,k}=\mathrm{Pool}_K(\mathbf{g}_i)_k$) and aligned to teacher features. $\mathcal{L}_{\text{LM}}$ backpropagates through the frozen decoder to adapt routing without altering the compute graph. To prioritize audio-dependent samples, a frozen probe scores the audio utility gap $\delta_i=\mathcal{L}_{\text{text-only}}(x_i,y_i)-\mathcal{L}_{\text{audio}}(x_i,a_i,y_i)$ to derive instance weights $w_i=\mathrm{clip}(1+\beta\delta_i,\,w_{\min},\,w_{\max})$, yielding
\begin{equation}
    \mathcal{L}_{\text{MoE}} = \sum_i w_i\,\mathcal{L}_{\text{CE}}(x_i,a_i,y_i) + \lambda\,\mathcal{L}_{\text{REPA}}.
\end{equation}
We freeze layer-wise routing indices from a reference pass to prevent expert drift. Since alignment is baked into weights, inference requires standard single-forward decoding—with zero extra tokens, multi-pass overhead, or routing metadata.


%% file: sections/4_Experiments.tex
\section{Experiments}

\subsection{Experimental Settings}

We implement all experiments using \texttt{ms-swift} and Megatron-LM~\citep{shoeybi2020megatronlmtrainingmultibillionparameter} on eight 80GB NVIDIA GPUs. The training pipeline utilizes \texttt{bf16} precision, FlashAttention, activation recomputation, and a 16{,}384-token sequence limit. We evaluate this setup across two backbones: the dense Qwen2.5-Omni-7B (with tensor parallelism) and the MoE Qwen3-Omni-30B-A3B (with tensor and expert parallelism), both utilizing audio-conditioned thinking dialogues. Appendix~\ref{app:training-hparams} details the hyperparameters and data configurations.

\subsection{Main Results}

\begin{table}[htbp]
  \centering
  \small
  \caption{Overall accuracy (\%) on the \textsc{UniVerseBench} across aggregated, language, and skill categories. 
  Language strata include \textit{LoC (Languages of China)}, \textit{ko} (Korean), and \textit{L-oth.} (other languages). 
  Skill strata encompass \textit{FE} (feature extraction), \textit{CR} (causal reasoning), \textit{PR} (pattern recognition), 
  \textit{CA} (comparison analysis), and \textit{ED} (error detection). 
  ``Post-trained'' denotes the best performance of models fine-tuned on our thinking-format post-training data.
  }
  \label{tab:overall_results}
  \renewcommand{\arraystretch}{1.2}
  \setlength{\tabcolsep}{4pt}
  \begin{tabular}{@{}l*{9}{c}@{}}
    \toprule
    \textit{Model} & \textit{Overall} & \textit{LoC*} & \textit{ko} & \textit{L-oth.} & \textit{FE} & \textit{CR} & \textit{PR} & \textit{CA} & \textit{ED} \\
    \midrule
    \textit{Kimi-audio} & 46.5 & 37.8 & 57.8 & 46.2 & 44.2 & 47.7 & 47.3 & 62.6 & 50.0 \\
    \textit{MidashengLM} & 40.9 & 36.4 & 46.5 & 40.9 & 39.5 & 41.6 & 39.9 & 54.4 & 31.9 \\
    \textit{Music Flamingo} & 54.4 & 47.3 & 61.6 & 55.1 & 54.1 & 57.1 & 53.6 & 51.8 & 55.3 \\
    \textit{Gemini 3 Flash} & 70.8 & 67.6 & \textit{\textbf{77.6}} & 69.5 & 68.7 & 73.3 & 71.3 & \textit{\textbf{84.9}} & 55.6 \\
    \textit{Qwen3.5-Omni-plus} & \textit{\textbf{74.4}} & \textit{\textbf{71.0}} & 75.5 & \textit{\textbf{75.7}} & \textit{\textbf{73.6}} & \textit{\textbf{80.1}} & \textit{\textbf{72.6}} & 71.4 & \textit{\textbf{77.1}} \\
    \midrule
    \textit{Qwen2.5-Omni-7B-Instruct} & 33.9 & 29.2 & 45.2 & 32.0 & 32.5 & 38.3 & 35.7 & 30.0 & 37.1 \\
    \textit{Qwen2.5-Omni-7B-Instruct (Post-trained)} & 48.8 & 44.4 & 57.6 & 47.8 & 47.4 & 52.4 & 54.4 & 45.7 & 48.6 \\
    \textit{Qwen3-Omni-30B-A3B} & 47.5 & 44.3 & 52.6 & 47.3 & 44.9 & 58.3 & 51.4 & 41.4 & 51.4 \\
    \textit{Qwen3-Omni-30B-A3B (Post-trained)} & 53.4 & 46.4 & 62.6 & 53.5 & 50.8 & 61.4 & 55.1 & 50.0 & 65.7 \\
    \bottomrule
  \end{tabular}
\end{table}



\subsubsection{Overall Performance Comparison}

Table~\ref{tab:overall_results} presents the overall accuracy on the \textsc{UniVerseBench}. Open-resource baselines---Kimi-audio~\citep{kimiteam2025kimiaudiotechnicalreport}, MidashengLM~\citep{dinkelMiDashengLMEfficientAudio2025}, Music Flamingo~\citep{ghosh2025musicflamingoscalingmusic}, Commercial Gemini 3 Flash, and Qwen3.5-Omni-Plus~\citep{qwenteam2026qwen35omnitechnicalreport}---are evaluated via direct option generation. In contrast, Qwen2.5-Omni and Qwen3-Omni are strictly evaluated in thinking mode, requiring both a valid reasoning chain and the correct final answer to demonstrate genuine understanding acquired through post-training.

Qwen3.5-Omni-Plus achieves the highest overall accuracy ($74.4\%$), serving as our upper-bound reference. 
Most models follow a consistent difficulty hierarchy: $\textit{LoC} > \textit{L-oth.} > \textit{ko}$, which is likely driven by the greater linguistic complexity and more heterogeneous audio sources inherent in the LoC stratum. Across skill strata, Qwen3.5-Omni-Plus leads in feature extraction (\textit{FE}), causal reasoning (\textit{CR}), pattern recognition (\textit{PR}), and error detection (\textit{ED}), whereas Gemini~3~Flash excels in comparison analysis (\textit{CA}; $74.8\%$). Crucially, thinking-format SFT consistently boosts performance under identical evaluation protocols, improving Qwen2.5-Omni by $14.9$ absolute percentage points (to $48.8\%$) and Qwen3-Omni by $5.9$ points (to $53.4\%$).

\subsubsection{Imbalanced Learning Strategies}

Table~\ref{tab:benchmark-ablation} compares the efficacy of data reweighting, preference optimization, and representation-aligned latent or encoder-side objectives on both Qwen backbones.

For \textbf{Qwen2.5-Omni}, language reweighting ($41.8\%$) and its combination with Text DPO ($42.6\%$) yield only modest gains over standard thinking-format SFT ($40.6\%$). Single-phase latent reasoning (Phase a) reaches $42.9\%$, while recurrent latent alignment (Phase b) achieves the peak performance of $48.8\%$ ($+8.2$ points over default SFT), establishing it as the optimal dense checkpoint. Audio DPO reaches $45.7\%$, outperforming early-stage latent reasoning but trailing Phase b.

For \textbf{Qwen3-Omni}, thinking-format SFT achieves the highest accuracy ($53.4\%$). Conversely, language reweighting ($51.9\%$) and Text DPO ($50.9\%$) degrade performance. Encoder-side REPA ($52.9\%$) and its combination with router-gate $\Delta\text{CE}$ weighting ($52.9\%$) provide stable adaptation, remaining competitive with SFT while preserving single-forward serving efficiency.

\newcommand{\qacc}[2]{%
  \makebox[2.8em][r]{#1}%
  \,${}\mid{}$\,%
  \makebox[2.8em][l]{#2}%
}

\begin{table}[htbp]
  \centering
  \small
  \caption{Ablation study of imbalanced learning and alignment strategies on the \textsc{UniVerseBench}. 
  Values report strict parseable accuracy (\%) formatted as \mbox{Qwen2.5-Omni} $\mid$ \mbox{Qwen3-Omni}. 
  Underlined entries denote the best performance for each backbone.
  }
  \label{tab:benchmark-ablation}
  \begin{tabular}{@{}lcc|lcc@{}}
    \toprule
    Method & Qwen2.5 & Qwen3 & Method & Qwen2.5 & Qwen3 \\
    \midrule
    SFT (w\textbackslash{} think) & 40.6 & \underline{53.4} & Audio DPO / Enc.-side REPA & 45.7 & 52.9 \\
    Lang.\ Loss & 41.8 & 51.9 & LR. + REPA (Phase a) & 42.9 & --- \\
    Lang.\ Loss + Text DPO & 42.6 & 50.9 & LR. + REPA (Phase b) / Route-b $\Delta$CE & \underline{48.8} & 52.9 \\
    \midrule
  \end{tabular}
\end{table}

\subsubsection{Correlation Analysis}
In Table~\ref{tab:train-count-correlation}, we compute Pearson's $r$ and Spearman's $\rho$ between post-training utterance counts ($n_\ell$) and per-language accuracy across $N{=}25$ languages ($n\ge10$) on the full benchmark, as well as across four primary skills. Across all strategies and backbones, no statistically significant positive correlation emerges ($\alpha{=}0.05$). Overall correlation coefficients remain near zero for both models. While weak negative correlations occasionally appear within specific skills (e.g., Spearman's $\rho$ on CA), overall accuracy remains fundamentally decoupled from training volume. 

These findings underscore two key conclusions:
(1) \textbf{Benchmark characteristics:} \textsc{UniVerseBench} evaluates culturally grounded musical understanding rather than text-language frequency, ensuring low-resource languages are not systematically disadvantaged.
(2) \textbf{Post-training dynamics:} Imbalanced-learning interventions enhance reasoning by reshaping internal representations, rather than merely amplifying high-frequency training data.

\begin{table*}[htbp]
  \centering
  \scriptsize
  \setlength{\tabcolsep}{2.8pt}
  \caption{Correlation between post-training utterance count and per-language adversarial accuracy ($N{=}25$; $n\geq10$).
  LR.\ represents Latent Reasoning. REPA++ represents REPA+$\Delta$CE+gates.
  Comparison Analysis merges plain + neutral variants.
  Category columns report Pearson's $r$ and Spearman's $\rho$ within each skill.}
  \label{tab:train-count-correlation}
  \begin{tabular}{@{}l rr rr  rr rr rr rr@{}}
    \toprule
    & \multicolumn{4}{c}{Overall} & \multicolumn{2}{c}{FE} & \multicolumn{2}{c}{CR} & \multicolumn{2}{c}{PR} & \multicolumn{2}{c}{CA} \\
    \cmidrule(lr){2-5} \cmidrule(lr){6-7} \cmidrule(lr){8-9} \cmidrule(lr){10-11} \cmidrule(lr){12-13}
    Strategy & $r$ & $p$ & $\rho$ & $p$ & $r$ & $\rho$ & $r$ & $\rho$ & $r$ & $\rho$ & $r$ & $\rho$ \\
    \midrule
    Q2.5 SFT (think) & 0.255 & 0.219 & 0.028 & 0.895 & 0.279 & 0.105 & 0.346 & 0.096 & $-$0.014 & 0.204 & $-$0.272 & $-$0.531 \\
    Q2.5 Lang.\ Loss & 0.270 & 0.193 & 0.095 & 0.651 & 0.350 & 0.140 & 0.099 & $-$0.077 & 0.062 & 0.031 & $-$0.130 & $-$0.480 \\
    Q2.5 Text DPO & 0.339 & 0.097 & 0.064 & 0.760 & 0.327 & $-$0.108 & 0.150 & 0.017 & 0.105 & 0.164 & 0.252 & $-$0.133 \\
    Q2.5 Audio DPO & 0.366 & 0.072 & 0.130 & 0.536 & 0.355 & 0.051 & 0.222 & 0.103 & 0.209 & 0.312 & $-$0.205 & $-$0.186 \\
    Q2.5 LR Phase-a & 0.312 & 0.129 & $-$0.037 & 0.862 & 0.353 & $-$0.121 & 0.207 & 0.113 & 0.049 & 0.211 & $-$0.158 & $-$0.442 \\
    Q2.5 LR Phase-b & 0.124 & 0.556 & $-$0.173 & 0.407 & 0.146 & $-$0.157 & 0.018 & $-$0.081 & 0.197 & 0.250 & $-$0.157 & $-$0.313 \\
    \midrule
    Q3 SFT (think) & 0.091 & 0.666 & 0.046 & 0.827 & 0.173 & 0.038 & $-$0.004 & 0.044 & $-$0.166 & $-$0.083 & 0.040 & $-$0.240 \\
    Q3 Lang.\ Loss & 0.038 & 0.858 & $-$0.133 & 0.525 & $-$0.001 & $-$0.202 & 0.090 & 0.037 & 0.016 & $-$0.042 & 0.056 & $-$0.204 \\
    Q3 Text DPO & 0.059 & 0.779 & $-$0.081 & 0.701 & 0.193 & $-$0.073 & $-$0.169 & $-$0.249 & $-$0.187 & 0.050 & $-$0.003 & $-$0.547 \\
    Q3 Enc.-side REPA & 0.123 & 0.559 & 0.120 & 0.568 & 0.199 & 0.125 & $-$0.204 & $-$0.090 & 0.165 & $-$0.009 & 0.150 & $-$0.238 \\
    Q3 Enc.-side REPA++ & 0.139 & 0.506 & 0.109 & 0.604 & 0.308 & 0.280 & $-$0.299 & $-$0.260 & $-$0.071 & $-$0.074 & $-$0.119 & $-$0.308 \\
    \bottomrule
  \end{tabular}
\end{table*}

\subsection{Cross-cultural Generalization}
We investigate whether post-training gains transfer across three fine-grained cultural axes: language exposure, Chinese regions, and Korean geographical provinces and melodic modes (\textit{tori}). Unless otherwise specified, evaluations utilize the optimal checkpoint for each backbone.

\paragraph{Language Transfer.}
Categorizing languages into four volume-versus-gain quadrants (Table~\ref{tab:all-languages-by-train}) reveals a clear decoupling between training scale and tuning efficacy. In the \textit{low-resource/high-gain} quadrant, Qwen2.5-Omni achieves substantial zero- and few-shot improvements on minimally exposed languages (e.g., Romanian $12.2\%\!\rightarrow\!58.1\%$; Danish $25.3\%\!\rightarrow\!50.7\%$). Conversely, the \textit{high-resource/low-gain} quadrant exhibits diminishing returns, with high-volume languages yielding marginal gains and occasionally regressing under Qwen3-Omni (e.g., Greek $68.3\%\!\rightarrow\!56.7\%$; Serbian $54.9\%\!\rightarrow\!43.7\%$). Thus, global post-training is not strictly Pareto-improving: latent reasoning on dense models effectively transfers structural musical priors to elevate weak baselines, whereas default SFT on MoE risks compromising previously robust representations on well-learned languages.

\begin{figure}[t]
  \centering
  \includegraphics[width=\linewidth]{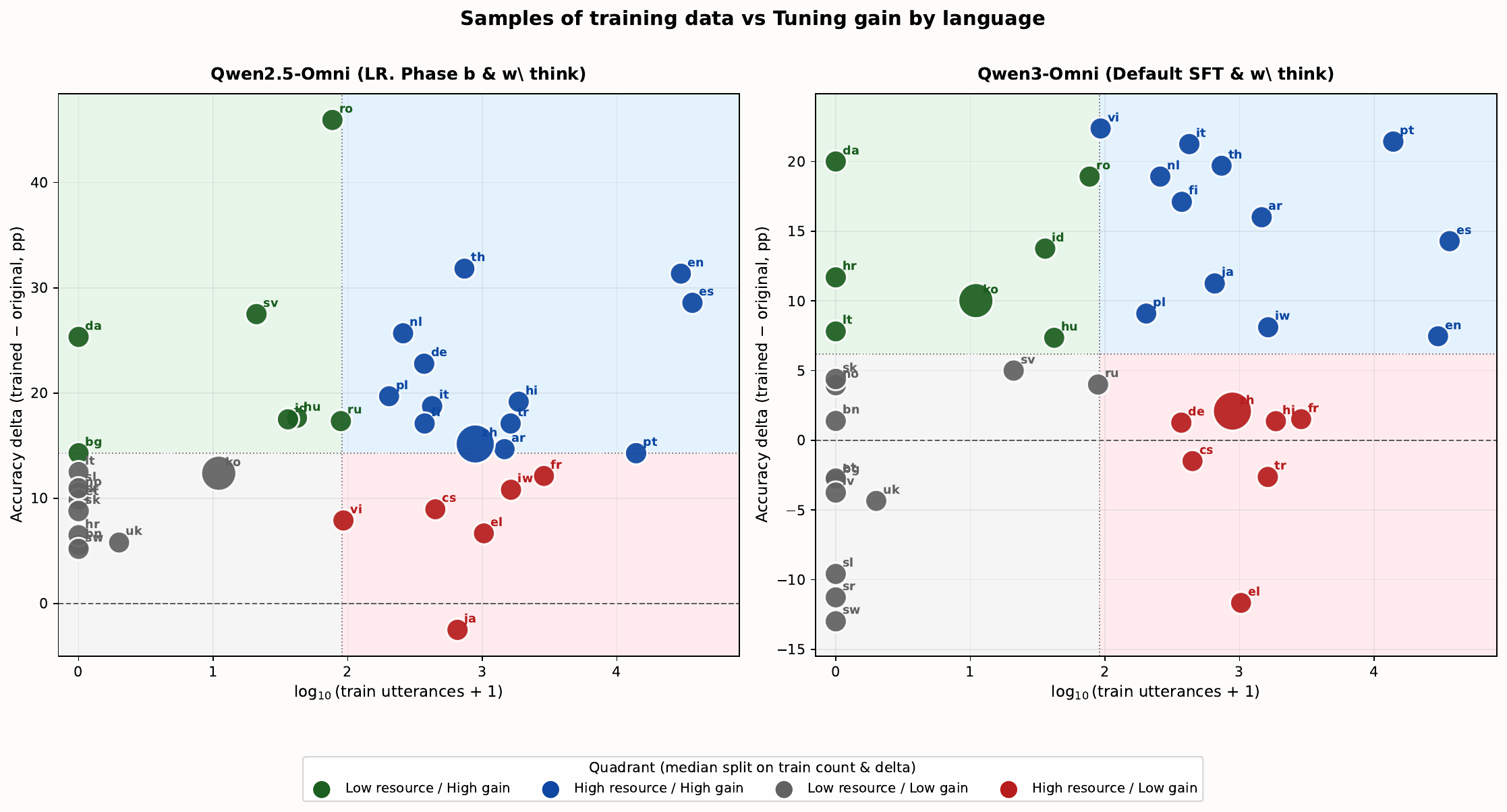}
  \caption{Per-language tuning gain versus training volume ($\log_{10}(n{+}1)$). Point size scales with sample size; quadrant axes indicate median splits on training volume and accuracy gain.}
  \label{fig:language_train_vs_gain_scatter}
\end{figure}

\paragraph{Chinese Regional Variation.}
Aggregating national data obscures significant geographic heterogeneity (Table~\ref{tab:chinese-region-tuning}). Qwen2.5-Omni markedly improves performance across initially weak regions (e.g., Jiangsu $34.4\%\!\rightarrow\!54.1\%$; Shanghai $34.5\%\!\rightarrow\!56.4\%$), though Qinghai and Hainan remain challenging. Qwen3-Omni achieves peak regional accuracy in areas like Chongqing ($75.0\%$) and Beijing ($63.6\%$), but regresses in regions where the base model was already strong (e.g., Jiangsu $59.0\%\!\rightarrow\!49.2\%$). This confirms that culturally grounded failure modes are acutely sensitive to intra-language regional variations.

\begin{table}[t]
  \centering
  \small
  \setlength{\tabcolsep}{8pt}
  \caption{Best post-training performance vs. original performance of Qwen2.5-Omni and Qwen3-Omni on the Chinese regional subset (sorted by mean trained accuracy). Guangdong* includes Hong Kong and Macau. Regions printed in \textcolor{violet}{violet} indicate those with official regional and indigenous languages. Format: original~$|$~trained.}
  \label{tab:chinese-region-tuning}
  \begin{tabular}{@{}llcllc@{}}
    \toprule
    Region & Qwen2.5-Omni & Qwen3-Omni & Region & Qwen2.5-Omni & Qwen3-Omni \\
    \midrule
    Chongqing & 31.2 | 75.0 & 62.5 | 75.0 & Ningxia & 8.7 | 43.5 & 56.5 | 43.5 \\
    Hebei & 31.0 | 56.9 & 37.9 | 48.3 & \textcolor{violet}{Guizhou} & 31.1 | 42.6 & 39.3 | 42.6 \\
    Shanghai & 34.5 | 56.4 & 40.0 | 47.3 & \textcolor{violet}{Taiwan} & 29.1 | 41.8 & 41.8 | 48.1 \\
    Liaoning & 18.5 | 55.6 & 55.6 | 59.3 & Shandong & 26.1 | 41.3 & 52.2 | 52.2 \\
    Jiangsu & 34.4 | 54.1 & 59.0 | 49.2 & Gansu & 28.3 | 41.3 & 30.4 | 43.5 \\
    Henan & 39.1 | 52.2 & 52.2 | 52.2 & Jiangxi & 26.2 | 41.0 & 47.5 | 50.8 \\
    \textcolor{violet}{Inner Mongolia} & 35.0 | 50.0 & 30.0 | 48.3 & Shaanxi & 18.2 | 40.9 & 56.8 | 47.7 \\
    \textcolor{violet}{Heilongjiang} & 27.3 | 50.0 & 27.3 | 45.5 & Anhui & 39.0 | 40.7 & 50.8 | 45.8 \\
    \textcolor{violet}{Tibet} & 40.7 | 48.1 & 51.9 | 55.6 & Tianjin & 33.3 | 40.0 & 53.3 | 60.0 \\
    \textcolor{violet}{Xinjiang} & 30.4 | 47.8 & 53.6 | 44.9 & Jilin & 32.3 | 38.7 & 32.3 | 35.5 \\
    Hubei & 31.2 | 46.9 & 35.9 | 45.3 & Guangdong*  & 18.5 | 38.5 & 27.7 | 38.5 \\
    \textcolor{violet}{Yunnan} & 30.2 | 46.5 & 27.9 | 44.2 & Sichuan & 10.3 | 33.3 & 56.4 | 48.7 \\
    Beijing & 18.2 | 45.5 & 72.7 | 63.6 & Shanxi & 40.9 | 29.5 & 47.7 | 38.6 \\
    Zhejiang & 33.3 | 45.2 & 54.8 | 50.0 & \textcolor{violet}{Hainan} & 16.7 | 26.7 & 40.0 | 33.3 \\
    Hunan & 40.5 | 45.2 & 42.9 | 35.7 & Qinghai & 38.6 | 26.3 & 42.1 | 45.6 \\
    Fujian & 26.5 | 44.1 & 32.4 | 44.1 & \textcolor{violet}{Guangxi} & 26.8 | 43.9 & 31.7 | 43.9 \\
    \bottomrule
  \end{tabular}
\end{table}

\paragraph{Korean Geography and Melodic Mode.}
Table~\ref{tab:korean-region-tori-best-tuned} evaluates Korean audio across provinces and melodic modes (\textit{tori}). Both backbones exhibit broad cross-regional gains (e.g., Pyeongan under Qwen2.5-Omni: $30.6\%\!\rightarrow\!62.9\%$; under Qwen3-Omni: $54.8\%\!\rightarrow\!71.0\%$). Stratifying by \textit{tori} reveals across-the-board improvements, particularly in \textit{Susimga} and \textit{Nanbongga}. Nevertheless, specific music-theoretic bottlenecks persist (e.g., \textit{Sin-gyeong} under Qwen3-Omni remains stagnant at $50.0\%$).

\begin{table}[htbp]
  \centering
  \small
  \setlength{\tabcolsep}{4pt}
  \caption{Korean accuracy stratified by region (left) and melodic mode (\textit{tori}, right). Format: original~$|$~trained.}
  \label{tab:korean-region-tori-best-tuned}
  \begin{tabular}{@{}llc|llc@{}}
    \toprule
    Region & Qwen2.5-Omni & Qwen3-Omni & Mode & Qwen2.5-Omni & Qwen3-Omni \\
    \midrule
    Gyeonggi & 41.5 | 51.8 & 52.3 | 52.3 & Yukjabaegi & 50.2 | 56.5 & 55.9 | 66.1 \\
    Gangwon & 58.7 | 61.9 & 50.8 | 60.3 & Menari & 46.7 | 54.4 & 50.9 | 56.2 \\
    Gyeongsang & 39.7 | 53.8 & 47.4 | 60.3 & Susimga & 37.1 | 61.0 & 40.0 | 65.7 \\
    Jeolla & 50.0 | 56.2 & 56.0 | 65.6 & Jin-gyeong & 42.2 | 54.1 & 55.0 | 55.0 \\
    Jeju & 44.4 | 59.3 & 54.3 | 59.3 & Gyeong & 28.6 | 71.4 & 57.1 | 64.3 \\
    Hwanghae & 45.2 | 65.8 & 50.0 | 69.2 & Ban-gyeong & 50.0 | 55.3 & 46.1 | 57.9 \\
    Pyeongan & 30.6 | 62.9 & 54.8 | 71.0 & Nanbongga & 40.0 | 65.0 & 57.5 | 73.8 \\
    Hamgyeong & 34.4 | 56.2 & 40.6 | 62.5 & Sin-gyeong & 30.8 | 53.8 & 30.8 | 50.0 \\
    Gyeonggi/Chungcheong & 36.4 | 54.5 & 36.4 | 63.6 & Jeju scale & 46.6 | 63.8 & 55.2 | 60.3 \\
    \bottomrule
  \end{tabular}
\end{table}

To sum up, 
Qwen2.5-Omni's latent-reasoning post-training preferentially elevates weak baselines, whereas Qwen3-Omni's default SFT maintains higher absolute accuracy but induces trade-offs that can degrade strong base languages. Therefore, achieving cross-cultural robustness demands evaluating granular geographic and music-theoretic strata—rather than relying solely on language-level averages—and selecting training objectives with explicit awareness of these cross-strata trade-offs. More details are discussed in Appendix~\ref{app:melodic-contour-case-study}.

%% file: sections/5_Conclusion.tex
\section{Conclusion and Future Work}

We introduced \textsc{UniVerse}, a reproducible solution that enhances the low-resource music understanding of diverse traditions. It consists of \textsc{UniVerseBench} and \textsc{UniVerseSet}, where one is a multilingual and music-oriented benchmark spanning over 38 languages, designed to evaluate the limitations of LALMs in culturally grounded folk-music understanding, and the other is a training dataset upon which multiple post-training strategies are investigated. By assessing post-training strategies on Qwen2.5-Omni and Qwen3-Omni, we challenge a central premise in the field: neither language-frequency exposure nor a unified global post-training recipe reliably guarantees competence in culture-specific musical reasoning.

While our semi-automated human--AI curation pipeline accelerates data synthesis, LLM-driven generation frequently struggles with expert constraints, underscoring the indispensable role of rigorous human verification for high-stakes evaluation. Empirically, three key insights emerge from our evaluations:

\textbf{First, training volume does not predict performance.} Across 25 post-training languages, utterance counts exhibit no significant correlation with model efficacy. \textsc{UniVerseBench} evaluates deep cultural grounding rather than mere text-language frequency, allowing low-resource languages to frequently outperform their high-resource counterparts.

\textbf{Second, cross-cultural transfer is highly stratified and operates across multiple scales.} Global post-training does not yield strict Pareto improvements. While latent reasoning in dense models consistently elevates weak baselines, full-parameter SFT in MoE architectures can inadvertently degrade performance on strong base languages. Substantial variance persists within fine-grained regional traditions (e.g., Chinese sub-regions) and music-theoretic dimensions (e.g., Korean \textit{tori}), demonstrating that language-level averages mask critical geographic vulnerabilities.

\textbf{Third, imbalanced-learning objectives act via skill- and modality-specific mechanisms.} While simple language reweighting yields negligible gains, representation-aligned interventions—such as recurrent latent alignment and encoder-side REPA—reshape internal reasoning and preserve serving efficiency without relying on proportional volume expansion.

To realize culturally faithful AI generation with a better interactive creation process, we prioritize three future directions: (1) collaborative platforms to convert expert knowledge into auditable alignment data; (2) controllable generation constrained by structural and regional musical priors; and (3) interactive co-creation workflows paired with dynamic benchmarks to comprehensively assess real-world human--AI collaboration.

%% file: sections/Appendix.tex
\newpage
\appendix
\section*{Appendix}
\addcontentsline{toc}{section}{Appendix}  
\setcounter{section}{0}
\renewcommand{\thesection}{\Alph{section}}
\renewcommand{\thesubsection}{\thesection.\arabic{subsection}}
\section{UniVerse Benchmark Verification}\label{app:universe_benchamrk_verification}

To ensure the precision, musical authenticity, and logical consistency of the evaluation items, we establish a systematic, multi-stage hybrid verification protocol that integrates automated filtering, expert annotation, AI-assisted modification, adversarial selection, and a final manual audit into a continuous workflow.

First, an automated quality and adversarial filtering pass is conducted on the raw pool of generated QA candidates to select a refined subset of 1,873 high-quality candidate QA pairs reserved for subsequent human validation. This process integrates baseline-agnostic quality screening with a programmatic adversarial selection pipeline designed to maximize evaluation difficulty. It prioritizes boosted questions, specifically those where baseline models failed or where Gemini optimization flagged misleading acoustic cues. To guarantee uniform benchmark density across the dataset, we enforce a strict minimum of five questions per audio clip, dynamically backfilling the quota using the baseline models' incorrect items.

Next, a multi-category human expert annotation phase is executed, where each of the 1,873 selected candidate items is assigned to at least one ethnomusicology or music theory expert. The experts evaluate the audio-text alignment and categorize each QA pair into one of three statuses:
\begin{itemize}[leftmargin=1.5em]
    \item \textbf{Directly Usable}: The question, options, and ground-truth answers are correct and require no modifications.
    \item \textbf{Requires Modification}: The item contains minor flaws such as ambiguous question phrasing, overlapping distractors, or slightly inaccurate musical terminology, and the expert provides explicit, granular feedback detailing the necessary corrections to the question stem or options.
    \item \textbf{Unusable}: The item suffers from fundamental conceptual or alignment errors.
\end{itemize}
To maintain the targeted benchmark scale of 1,873 items, unusable pairs trigger a rejection-replacement process where the expert reviews alternative candidate QA pairs generated from the same audio clip that were not initially selected, aiming in the worst-case scenario to find an alternative pair that can be rendered usable through minor modifications.

Following the human annotation phase, we leverage DeepSeek-4.0-Pro for secondary quality control. DeepSeek-4.0-Pro consolidates the expert annotations, textual modification guidelines, and replacement candidates to reconstruct and refine the question stems, options, and annotations. After processing these updates, the model conducts a secondary automated check over the entire consolidated dataset to verify formatting uniformity and logical consistency before exporting.

Finally, a senior panel of musicology experts conducts a final manual audit of the compiled dataset. This final pass manually verifies that all human-requested modifications and replacement pairs were correctly integrated and that no secondary errors were introduced, establishing the final UniVerse benchmark.

\subsection{Data Collection and Alignment}\label{app:data_collection_details}
To ensure the cultural diversity of our low-resource music corpus, our metadata curation specifically targeted traditional and regional music. We queried a global music streaming service using a carefully compiled list of \textbf{770 traditional and regional genre tags}. 
Once the track titles and corresponding artist names were extracted, we utilized the SoundCharts platform\footnote{https://soundcharts.com/} to bridge the textual metadata with raw audio. By querying these title-artist pairs via SoundCharts, we located their linked YouTube identifiers, which were subsequently used to download raw audio.

\subsection{Cross-Validation Heuristics}\label{app:feature_labeling_details}
Since the feature annotations (scores, lyrics, captions) are generated by independent models, we employed Qwen3-Next-80B-A3B-Instruct to conduct rigorous cross-modal verification. A track is excluded from the final training corpus if the LLM identifies any of the following fatal inconsistencies:
\begin{itemize}[leftmargin=1.0em]
    \item \textbf{Symbolic-Acoustic Mismatch:} Conflicts in musical attributes, such as key signatures or time signatures, between the ABC notation (from SheetSage) and the acoustic captions (from Qwen3-Omni-Captioner).
    \item \textbf{Linguistic Hallucination:} Instances where the caption explicitly quotes or references specific lyrics that are entirely absent from the Qwen3-ASR transcription.
\end{itemize}

\subsection{Dialogue Synthesis Rules and Constraints}\label{app:dialogue_synthesis_details}
To simulate realistic, multi-turn listening sessions while preventing data leakage, we enforced a strict set of rule-based constraints during the LLM dialogue generation.

\subsubsection{User Profile and Task Sampling}
Before generating a dialogue, a structured user profile is instantiated for each track:
\begin{itemize}[leftmargin=1.0em]
    \item \textbf{Language Alignment:} The conversation language is dynamically matched to the lyric metadata. If the language field is missing, empty, or labeled as unknown, the pipeline robustly defaults to English.
    \item \textbf{Conditional Persona Mapping:} To prevent logical contradictions, user personas are generated via a two-stage sequential sampling process. We first select a demographic style group (\textit{Gen Z, Middle-aged, Elderly,} or \textit{Professional Academic}). We then sample a music expertise level restricted to that demographic's permitted subset. For instance, a Gen Z profile is restricted to \textit{Lay listener} or \textit{Amateur musician}, whereas a Professional Academic is restricted to \textit{Professional composer} or \textit{Cultural historian}.
    \item \textbf{Turn and Intent Constraints:} The dialogue length $T$ is randomized between 1 and 9 turns. We sample $I$ unique intents ($1 \le I \le \min(|Pool|, T)$) from a pool of 10 tasks. These include six general inquiries (\textit{Factual Inquiry, Structural Analysis, Subjective Appreciation, Creative/Cultural Inquiry, Technical Verification,} and \textit{Ethnomusicological Comparison}) and four academic tasks (\textit{Multiple-Choice Question (MCQ), Fill-in-the-blanks, Short-answer Analysis,} and \textit{True/False Inquiry}).
    \item \textbf{Mode Assignment:} For academic tasks, we assign either a \textit{Direct Answer} mode (optimized for zero-shot quantitative benchmark evaluation) or an \textit{Analytical} mode (optimized for reasoning-heavy instruction tuning), with a 50\% probability each.
\end{itemize}

\subsubsection{Query Constraints for Blind Simulation}
Both the user and the assistant are framed as blind listeners with no prior access to the text annotations. User queries are heavily constrained to prevent answer leakage:
\begin{itemize}[leftmargin=1.0em]
    \item \textbf{Feature Anchoring:} Questions are strictly limited to the verified features (score, lyrics, or captions) actually available for that specific track.
    \item \textbf{Information Isolation:} Queries must be concise and free of contextual hints, descriptive adjectives, or lyric quotes that could reveal the ground-truth answer (e.g., prohibiting leading phrases like ``Since this is a G-minor song...'').
    \item \textbf{Task-Specific Formatting:} For fill-in-the-blank tasks, the user provides only the template sentence containing the blank placeholder. For MCQs, all options (A, B, C, D) are constrained to have similar lengths, vocabulary, and tone to prevent heuristic guessing.
\end{itemize}

\subsubsection{Assistant Reasoning and Output Rules}
To reduce factual hallucinations, the assistant's behavior is explicitly restricted in both its latent reasoning and final response:
\begin{itemize}[leftmargin=1.0em]
    \item \textbf{Verbatim Grounding:} Each assistant turn must feature a reference block citing the exact source keys (e.g., \texttt{caption}) and verbatim substrings from the extracted annotations to anchor deductions in verified facts.
    \item \textbf{Auditory Simulation:} The assistant's step-by-step thinking trace must be articulated as direct listening experiences (describing timbre, rhythm, and phonetic features) instead of explicitly mentioning the underlying JSON data sources.
    \item \textbf{Conditional Feedback Modes:} In \textit{Direct Answer Mode}, the final output must contain only the exact option or term (e.g., ``A'' or ``True'') without greetings. In \textit{Analytical Mode}, the assistant must provide a comprehensive, structured musicological explanation.
\end{itemize}

\begin{figure}[htbp]
    \centering
    \includegraphics[width=0.95\linewidth]{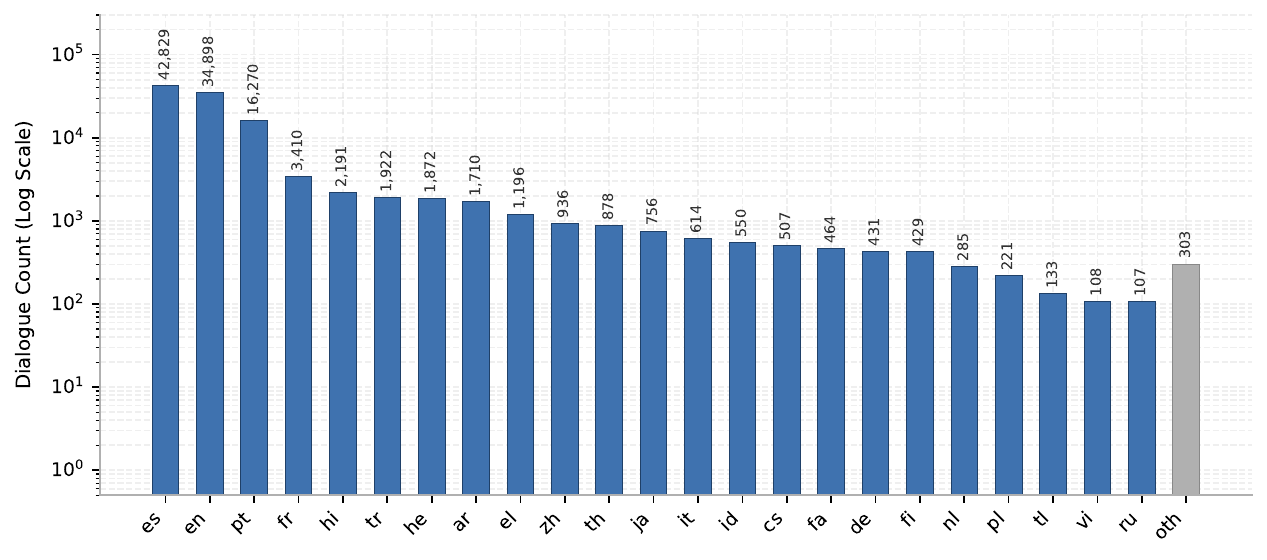}
    \caption{Language distribution of the synthesized corpus plotted on a logarithmic scale. The label \texttt{oth} aggregates all remaining low-frequency languages with fewer than 100 dialogues.}
    \label{fig:lang_dist}
\end{figure}

\section{Detailed Parameters of Different Training Strategies}\label{app:training-hparams}
All completed runs are trained with Megatron-LM through \texttt{ms-swift} on eight 80\,GB GPUs,
using FlashAttention, full activation recomputation, maximum sequence length $16384$,
one epoch, and a $1\%$ validation split.
All supervised and preference stages use the \textbf{full} post-training corpus of 
(thinking-format jsonl, or the corresponding full preference-pair export);
we do not use reduced-scale subsets.
Thinking-format data enable \texttt{add\_non\_thinking\_prefix}.
Vision towers remain frozen; audio generation is disabled during training.
\paragraph{Hyperparameters of Backbones.}
Qwen2.5-Omni is trained as a dense model with tensor parallel size $4$
(sequence parallel on, except Phase~b where it is disabled).
Qwen3-Omni uses tensor parallel $8$ and expert parallel $8$,
MoE auxiliary loss coefficient $10^{-6}$, and expert capacity factor $2.0$.
Both use CPU optimizer offload (fraction $0.5$; Text DPO on Qwen3-Omni uses $0.64$)
with the precision-aware optimizer.
Unless noted, SFT uses $\mathrm{lr}=10^{-5}$, audio-tower $\mathrm{vit\_lr}=10^{-6}$,
warmup fraction $0.05$, and $\mathrm{min\_lr}=10^{-6}$,
updating the full LLM together with \texttt{visual.audio\_tower}.
\paragraph{Qwen2.5-Omni Special Setting.}
SFT (w/ think) fine-tunes on the full thinking corpus with micro-/global batch sizes $4$/$8$.
Lang.\ Loss uses the same recipe with language-level \texttt{sample\_weight} on the full corpus.
Latent Reasoning Phase~a trains on the full lang-loss corpus
with batch sizes $4$/$16$, $K{=}6$ latent steps, and REPA weight $\lambda{=}0.5$;
Phase~b continues the Phase~a checkpoint with batch sizes $1$/$4$ under recurrent decoding
(same $K$ and $\lambda$).
Text DPO is initialized from the lang-loss SFT checkpoint and trains the full LLM plus audio tower
on full text-preference pairs with $\mathrm{lr}=10^{-7}$, $\mathrm{vit\_lr}=5{\times}10^{-7}$,
$\beta{=}0.03$, warmup $0.1$, $\mathrm{clip\_grad}{=}0.25$, and batch sizes $1$/$4$.
Vanilla Audio DPO freezes the LLM, trains only the audio tower on full M2 audio-swap pairs
with $\mathrm{lr}=\mathrm{vit\_lr}=10^{-6}$, $\beta{=}0.03$, warmup $0.1$,
$\mathrm{clip\_grad}{=}0.25$, and batch sizes $2$/$4$ (no chosen-likelihood anchor).
\paragraph{Qwen3-Omni Special Setting.}
SFT (w/ think) fine-tunes on the full thinking corpus with batch sizes $1$/$4$.
Lang.\ Loss matches this setup with language \texttt{sample\_weight} on the full corpus.
Encoder-side REPA initializes from the full-corpus think SFT checkpoint, freezes the MoE decoder,
and updates the audio tower with projections on the full lang-loss corpus
(batch sizes $2$/$32$, $K{=}6$, $\lambda_{\mathrm{REPA}}{=}0.1$, $\mathrm{vit\_lr}{=}10^{-5}$).
Encoder-side REPA~+~$\Delta$CE~+~gates uses the same initialization and batch sizes,
but optimizes audio parameters together with MoE router gates on the full $\Delta$CE-weighted corpus
($\mathrm{lr}=\mathrm{vit\_lr}{=}10^{-5}$).
Text DPO follows the Qwen2.5-Omni text-DPO schedule (full LLM~+~audio; $\beta{=}0.03$; batch $1$/$4$)
on full text-preference pairs, starting from the lang-loss SFT checkpoint.

\begin{table}[htbp]
  \centering
  \scriptsize
  \setlength{\tabcolsep}{10pt}
  \caption{Per-language accuracy (original $|$ trained) for all benchmark languages. Rows are grouped by the Qwen2.5-Omni training-count vs.\ gain quadrant (median splits on $\log_{10}(n{+}1)$ and $\Delta{=}\mathrm{trained}{-}\mathrm{original}$): green\,=\,low-resource/high-gain, blue\,=\,high-resource/high-gain, grey\,=\,low-resource/low-gain, red\,=\,high-resource/low-gain; within each block, sorted by $n$ (high$\to$low).}
  \label{tab:all-languages-by-train}
  \begin{tabular}{@{}llcllc@{}}
    \toprule
    Language [\#] & Qwen2.5-omni & Qwen3-omni & Language [\#] & Qwen2.5-omni & Qwen3-omni \\
    \midrule
    \cellcolor[HTML]{E8F5E9}{ru (Russian) [88]} & \cellcolor[HTML]{E8F5E9}{30.67 | 48.00} & \cellcolor[HTML]{F5F5F5}{50.67 | 54.67} & \cellcolor[HTML]{E8F5E9}{ro (Romanian) [76]} & \cellcolor[HTML]{E8F5E9}{12.16 | 58.11} & \cellcolor[HTML]{E8F5E9}{48.65 | 67.57} \\
    \cellcolor[HTML]{E8F5E9}{hu (Hungarian) [41]} & \cellcolor[HTML]{E8F5E9}{25.00 | 42.65} & \cellcolor[HTML]{E8F5E9}{39.71 | 47.06} & \cellcolor[HTML]{E8F5E9}{id (Indonesian) [35]} & \cellcolor[HTML]{E8F5E9}{37.50 | 55.00} & \cellcolor[HTML]{E8F5E9}{42.50 | 56.25} \\
    \cellcolor[HTML]{E8F5E9}{sv (Swedish) [20]} & \cellcolor[HTML]{E8F5E9}{23.75 | 51.25} & \cellcolor[HTML]{F5F5F5}{41.25 | 46.25} & \cellcolor[HTML]{E8F5E9}{bg (Bulgarian) [0]} & \cellcolor[HTML]{E8F5E9}{31.43 | 45.71} & \cellcolor[HTML]{F5F5F5}{52.86 | 50.00} \\
    \cellcolor[HTML]{E8F5E9}{da (Danish) [0]} & \cellcolor[HTML]{E8F5E9}{25.33 | 50.67} & \cellcolor[HTML]{E8F5E9}{36.00 | 56.00} & \cellcolor[HTML]{E3F2FD}{es (Spanish) [36476]} & \cellcolor[HTML]{E3F2FD}{25.97 | 54.55} & \cellcolor[HTML]{E3F2FD}{37.66 | 51.95} \\
    \cellcolor[HTML]{E3F2FD}{en (English) [29936]} & \cellcolor[HTML]{E3F2FD}{20.90 | 52.24} & \cellcolor[HTML]{E3F2FD}{58.21 | 65.67} & \cellcolor[HTML]{E3F2FD}{pt (Portuguese) [13911]} & \cellcolor[HTML]{E3F2FD}{33.33 | 47.62} & \cellcolor[HTML]{E3F2FD}{41.67 | 63.10} \\
    \cellcolor[HTML]{E3F2FD}{hi (Hindi) [1864]} & \cellcolor[HTML]{E3F2FD}{35.62 | 54.79} & \cellcolor[HTML]{FFEBEE}{60.27 | 61.64} & \cellcolor[HTML]{E3F2FD}{tr (Turkish) [1624]} & \cellcolor[HTML]{E3F2FD}{26.32 | 43.42} & \cellcolor[HTML]{FFEBEE}{52.63 | 50.00} \\
    \cellcolor[HTML]{E3F2FD}{ar (Arabic) [1463]} & \cellcolor[HTML]{E3F2FD}{34.67 | 49.33} & \cellcolor[HTML]{E3F2FD}{42.67 | 58.67} & \cellcolor[HTML]{E3F2FD}{zh (Chinese) [886]} & \cellcolor[HTML]{E3F2FD}{29.24 | 44.40} & \cellcolor[HTML]{FFEBEE}{44.26 | 46.35} \\
    \cellcolor[HTML]{E3F2FD}{th (Thai) [736]} & \cellcolor[HTML]{E3F2FD}{30.30 | 62.12} & \cellcolor[HTML]{E3F2FD}{54.55 | 74.24} & \cellcolor[HTML]{E3F2FD}{it (Italian) [423]} & \cellcolor[HTML]{E3F2FD}{31.25 | 50.00} & \cellcolor[HTML]{E3F2FD}{40.00 | 61.25} \\
    \cellcolor[HTML]{E3F2FD}{fi (Finnish) [372]} & \cellcolor[HTML]{E3F2FD}{32.89 | 50.00} & \cellcolor[HTML]{E3F2FD}{42.11 | 59.21} & \cellcolor[HTML]{E3F2FD}{de (German) [369]} & \cellcolor[HTML]{E3F2FD}{21.52 | 44.30} & \cellcolor[HTML]{FFEBEE}{50.63 | 51.90} \\
    \cellcolor[HTML]{E3F2FD}{nl (Dutch) [257]} & \cellcolor[HTML]{E3F2FD}{27.03 | 52.70} & \cellcolor[HTML]{E3F2FD}{29.73 | 48.65} & \cellcolor[HTML]{E3F2FD}{pl (Polish) [202]} & \cellcolor[HTML]{E3F2FD}{33.33 | 53.03} & \cellcolor[HTML]{E3F2FD}{53.03 | 62.12} \\
    \cellcolor[HTML]{F5F5F5}{ko (Korean) [10]} & \cellcolor[HTML]{F5F5F5}{45.19 | 57.56} & \cellcolor[HTML]{E8F5E9}{52.55 | 62.57} & \cellcolor[HTML]{F5F5F5}{uk (Ukrainian) [1]} & \cellcolor[HTML]{F5F5F5}{40.58 | 46.38} & \cellcolor[HTML]{F5F5F5}{56.52 | 52.17} \\
    \cellcolor[HTML]{F5F5F5}{bn (Bengali) [0]} & \cellcolor[HTML]{F5F5F5}{31.94 | 37.50} & \cellcolor[HTML]{F5F5F5}{36.11 | 37.50} & \cellcolor[HTML]{F5F5F5}{hr (Croatian) [0]} & \cellcolor[HTML]{F5F5F5}{32.47 | 38.96} & \cellcolor[HTML]{E8F5E9}{40.26 | 51.95} \\
    \cellcolor[HTML]{F5F5F5}{et (Estonian) [0]} & \cellcolor[HTML]{F5F5F5}{31.51 | 41.10} & \cellcolor[HTML]{F5F5F5}{49.32 | 46.58} & \cellcolor[HTML]{F5F5F5}{lv (Latvian) [0]} & \cellcolor[HTML]{F5F5F5}{38.75 | 48.75} & \cellcolor[HTML]{F5F5F5}{46.25 | 42.50} \\
    \cellcolor[HTML]{F5F5F5}{lt (Lithuanian) [0]} & \cellcolor[HTML]{F5F5F5}{28.12 | 40.62} & \cellcolor[HTML]{E8F5E9}{40.62 | 48.44} & \cellcolor[HTML]{F5F5F5}{no (Norwegian) [0]} & \cellcolor[HTML]{F5F5F5}{26.32 | 36.84} & \cellcolor[HTML]{F5F5F5}{50.00 | 53.95} \\
    \cellcolor[HTML]{F5F5F5}{sr (Serbian) [0]} & \cellcolor[HTML]{F5F5F5}{32.39 | 42.25} & \cellcolor[HTML]{F5F5F5}{54.93 | 43.66} & \cellcolor[HTML]{F5F5F5}{sk (Slovak) [0]} & \cellcolor[HTML]{F5F5F5}{31.87 | 40.66} & \cellcolor[HTML]{F5F5F5}{43.96 | 48.35} \\
    \cellcolor[HTML]{F5F5F5}{sl (Slovenian) [0]} & \cellcolor[HTML]{F5F5F5}{43.84 | 54.79} & \cellcolor[HTML]{F5F5F5}{60.27 | 50.68} & \cellcolor[HTML]{F5F5F5}{sw (Swahili) [0]} & \cellcolor[HTML]{F5F5F5}{40.26 | 45.45} & \cellcolor[HTML]{F5F5F5}{51.95 | 38.96} \\
    \cellcolor[HTML]{FFEBEE}{fr (French) [2874]} & \cellcolor[HTML]{FFEBEE}{39.39 | 51.52} & \cellcolor[HTML]{FFEBEE}{53.03 | 54.55} & \cellcolor[HTML]{FFEBEE}{iw (Hebrew) [1634]} & \cellcolor[HTML]{FFEBEE}{36.49 | 47.30} & \cellcolor[HTML]{E3F2FD}{33.78 | 41.89} \\
    \cellcolor[HTML]{FFEBEE}{el (Greek) [1027]} & \cellcolor[HTML]{FFEBEE}{45.00 | 51.67} & \cellcolor[HTML]{FFEBEE}{68.33 | 56.67} & \cellcolor[HTML]{FFEBEE}{ja (Japanese) [654]} & \cellcolor[HTML]{FFEBEE}{35.00 | 32.50} & \cellcolor[HTML]{E3F2FD}{43.75 | 55.00} \\
    \cellcolor[HTML]{FFEBEE}{cs (Czech) [447]} & \cellcolor[HTML]{FFEBEE}{37.31 | 46.27} & \cellcolor[HTML]{FFEBEE}{62.69 | 61.19} & \cellcolor[HTML]{FFEBEE}{vi (Vietnamese) [92]} & \cellcolor[HTML]{FFEBEE}{47.37 | 55.26} & \cellcolor[HTML]{E3F2FD}{38.16 | 60.53} \\
    \bottomrule
  \end{tabular}
\end{table}

\section{Case Study}
\begin{figure}[htb] 
    \centering 
    \includegraphics[width=\textwidth]{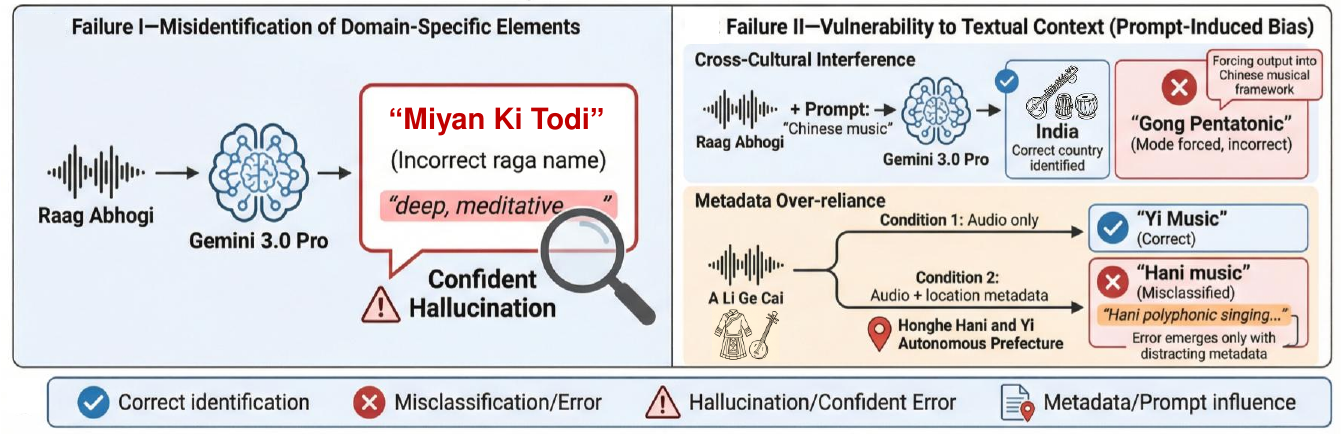} 
    \caption{Two systematic failure modes of LALMs on Raag Abhogi and A Li Ge Cai: misidentification \& prompt-induced bias.} 
    \label{fig:gemini_error} 
\end{figure}

\subsection{Keyword Concentration in Melodic-Contour Questions}
\label{app:melodic-contour-case-study}
We analyze melodic-contour questions to investigate pipeline limitations stemming from the overrepresentation of the label \texttt{undulating}. Specifically, we quantify this label bias, evaluate whether models exploit it as a shortcut, and trace its root cause across CantoCore, its automatic parser, and Gemini-based generation.

\subsubsection{Examination setup}
We restricted the audit to questions written in Chinese or English. 
Then, we programmatically and manually searched the benchmark for melodic-contour questions, which returned 191 questions, each paired with a distinct audio recording. Two examples from the retained subset are shown below.

\begin{center}
\begin{minipage}{0.92\linewidth}
\small
\hrule
\vspace{0.5em}
\textbf{Example 1.} Which of the following best describes the melodic contour of the vocal line in this excerpt?
\begin{enumerate}[label=\Alph*., leftmargin=2em, nosep]
    \item Subtle Segmented Descents
    \item Staccato
    \item Arpeggiated
    \item Undulating
\end{enumerate}
\textit{Correct answer: D. Undulating}

\vspace{0.6em}
\textbf{Example 2.} Based on what you hear, how would you characterize the melodic contour of this piece?
\begin{enumerate}[label=\Alph*., leftmargin=2em, nosep]
    \item Arched, undulating, and wave-like
    \item Linear and descending, with arched segments
    \item pointillist smooth
    \item syncopated jagged
\end{enumerate}
\textit{Correct answer: A. Arched, undulating, and wave-like}
\vspace{0.5em}
\hrule
\end{minipage}
\end{center}

We then extracted the melodic-contour feature produced by our CantoCore parser.
Table~\ref{tab:melodic-contour-audit} summarizes the audit results for this 191-question subset: it reports the proportion of correct options containing \texttt{Undulating} and, for the corresponding recordings, the distribution of parser-derived melodic-contour features that contain only \texttt{Undulating}, contain \texttt{Undulating} together with other labels, or do not contain \texttt{Undulating}. 
Note that \texttt{Undulating} occurs in 178/191 correct answers (93.19\%), this is a downstream answer regularity and therefore motivates the shortcut analysis below. We call these 178 questions undulating-positive questions in later sections.

\begin{table}[htbp]
  \caption{Answer-label concentration and CantoCore alignment.}
  \label{tab:melodic-contour-audit}
  \centering
  \small
  \begin{tabular}{@{}p{0.68\linewidth}rr@{}}
    \toprule
    Observation & Count & Proportion (\%) \\
    \midrule
    Correct option contains \texttt{Undulating} & 178/191 & 93.19 \\
    CantoCore contains only \texttt{Undulating} & 95/178 & 53.37 \\
    CantoCore is multi-coded and includes \texttt{Undulating} & 83/178 & 46.63 \\
    CantoCore does not include \texttt{Undulating} & 0/178 & 0.00 \\
    \bottomrule
  \end{tabular}
\end{table}


\subsubsection{Model results and the limits of the shortcut claim}
The concentration of the semantic label \texttt{Undulating} in every correct option creates a corpus-level cue that a model could potentially use without resolving the audio in full. 
Thus, we examine the results of every Qwen2.5-Omni and Qwen3-Omni configuration on 178 undulating-positive questions reported in Tables~\ref{tab:melodic-contour-model-summary}. Overall accuracy treats invalid outputs as incorrect, whereas valid-only accuracy removes them from the denominator. 

Full-response overall accuracy ranges from 41.01--54.49\% for Qwen2.5-Omni and 43.26--57.87\% for Qwen3; options-only accuracy ranges from 39.89--48.31\% and 47.75--65.17\%, respectively. 
 This heterogeneous and generally moderate performance does not exhibit a uniformly high accuracy that would support a claim of systematic shortcut use. 
The more direct finding of this case study is the answer regularity itself and what it reveals about the construction pipeline.



\begin{table}[t]
  \caption{Qwen2.5-Omni and Qwen3-Omni results on 178 Undulating-positive questions. "Valid" denotes valid response count ($N=178$). Accuracy is reported as Overall/Valid-only (\%) for Full response, and Overall (\%) for Options only. A dash denotes no evaluated setting.}
  \label{tab:melodic-contour-model-summary}
  \centering
  \scriptsize
  \setlength{\tabcolsep}{2pt} 
  \begin{minipage}[t]{0.48\textwidth}
    \centering
    \textbf{Qwen2.5-Omni}
    \vspace{2pt}
    \begin{tabular}{@{} >{\raggedright\arraybackslash}p{2.3cm} >{\centering\arraybackslash}p{0.8cm} >{\centering\arraybackslash}p{1.9cm} >{\centering\arraybackslash}p{1.1cm} @{}}
      \toprule
      & \multicolumn{2}{c}{Full response} & \multicolumn{1}{c}{Options only} \\
      \cmidrule(lr){2-3}\cmidrule(l){4-4}
      Configuration & Valid & Acc (\%) & Acc (\%) \\
      \midrule
      Original        & 176 & 44.38 / 44.89 & 48.31 \\
      Default SFT       & 163 & 41.01 / 44.79 & 42.13 \\
      Lang. Loss   & 166 & 44.38 / 47.59 & 42.70 \\
      Text DPO        & 168 & 53.93 / 57.14 & 42.13 \\
      Audio DPO & 161 & 42.70 / 47.20 & 39.89 \\
      LR Phase-a & 158 & 45.51 / 51.27 & --    \\
      LR Phase-b & 178 & 54.49 / 54.49 & 44.38 \\
      \bottomrule
    \end{tabular}
  \end{minipage}
  \hfill
  \begin{minipage}[t]{0.48\textwidth}
    \centering
    \textbf{Qwen3-Omni}
    \vspace{2pt}
    \begin{tabular}{@{} >{\raggedright\arraybackslash}p{2.3cm} >{\centering\arraybackslash}p{0.8cm} >{\centering\arraybackslash}p{1.9cm} >{\centering\arraybackslash}p{1.1cm} @{}}
      \toprule
      & \multicolumn{2}{c}{Full response} & \multicolumn{1}{c}{Options only} \\
      \cmidrule(lr){2-3}\cmidrule(l){4-4}
      Configuration & Valid & Acc (\%) & Acc (\%) \\
      \midrule
      Original          & 154 & 57.87 / 66.88 & 65.17 \\
      Default SFT         & 158 & 47.19 / 53.16 & 49.44 \\
      Lang. Loss     & 158 & 46.63 / 52.53 & 47.75 \\
      Text DPO          & 160 & 46.63 / 51.88 & 47.75 \\
      Enc.-side REPA        & 159 & 48.88 / 54.72 & 50.56 \\
      Enc.-side REPA++ & 158 & 43.26 / 48.73 & 49.44 \\
      \bottomrule
    \end{tabular}
  \end{minipage}
\end{table}

\subsubsection{Examining the automatic Cantocore parser}
The CantoCore specification treats melodic contour as a phrase-level qualitative character and explicitly requires annotator discretion. Temporary direction changes that do not materially alter the dominant contour---particularly changes lasting only one or two notes---should be ignored; otherwise, the specification warns, too many contours will be classified as \emph{undulating}, reducing the character's informativeness. 

Our CantoCore parser implementation provides a deterministic operationalization of this guidance.
The parser reads the human-transcribed XML score, segments phrases at rests or breath marks, and converts every remaining interval into an ascending or descending sign. A phrase is labeled \texttt{Undulating} whenever this sign sequence changes direction more than once; the recording-level output is the set union of all phrase labels.
The following examples illustrate this rule:

\begin{equation*}
\begin{aligned}
\boxed{C \nearrow D \searrow C \nearrow D}
&\quad\Longrightarrow\quad (+,-,+)
\quad\Longrightarrow\quad \texttt{Undulating},\\[0.4em]
\boxed{C \nearrow E \searrow D \searrow B \nearrow C}
&\quad\Longrightarrow\quad (+,-,-,+)
\quad\Longrightarrow\quad \texttt{Undulating}.
\end{aligned}
\end{equation*}

This operationalization is reproducible but has four limitations. First, reversals caused by ornaments, passing tones, or phrase-final motion may satisfy the symbolic rule without shaping the perceived principal contour. Second, one qualifying phrase adds \texttt{Undulating} regardless of its duration or frequency, so the recording-level union establishes presence rather than predominance. Third, even human-produced scores can simplify glides, microtonal motion, and culturally specific vocal inflections. Finally, rest- and breath-based boundaries need not coincide with perceived musical syntax, and different segmentation can change the number of within-phrase reversals. 

Thus, complete parser alignment for the 178 undulating-positive recordings demonstrates consistency with the automatic extraction rule, not universal agreement with listeners' holistic perception.
This is a trade-off for scalable, reproducible annotation and motivates targeted perceptual validation.
On the other hand, fully replacing this stage with human annotation under the CantoCore protocol would not remove all interpretive limitations, which will be discussed later, and would require considerably more time and labor. The relevant goal is therefore to understand and validate this trade-off, rather than to treat either automatic extraction or exhaustive human annotation as an unqualified reference.

\subsubsection{Granularity considerations at the Gemini question-generation stage}
Gemini faithfully transfers the supplied CantoCore label or complete multi-label set. The limitation instead arises from scope: the unordered recording-level set indicates that a contour occurs in at least one parser-defined phrase, but contains no location, duration, frequency, or prominence information. 
When Gemini converts this representation into a natural-language multiple-choice question, however, the wording often refers to the contour of the piece, song, vocal line, or main theme, and may use expressions such as \emph{overall}, \emph{primary}, or \emph{predominant}.
This can therefore expand local presence into a claim of global characterization.

Exact transfer guarantees traceability, but not alignment between question scope and evidence scope. A listener may recognize a local undulating phrase while perceiving another contour as dominant, so accuracy can conflate holistic audio judgment with recovery of a score-derived label. Global wording should therefore, if possible, be supported by richer phrase-level evidence or targeted perceptual review; otherwise, presence-based wording is more appropriate.

\subsubsection{Examining CantoCore as a human reference}
\label{discuss:cantocore_discuss}
CantoCore provides a common vocabulary for structured comparison while leaving room for subjective perceptual interpretation: its reported mean inter-rater reliability is $\kappa=0.47$, and $\kappa=0.37$ for melodic contour. Phrase grouping, multi-coding, and prominence therefore require judgment even under manual annotation, which is also substantially more costly. This perspective also clarifies our design choice. Score-based automation improves scalability, consistency, and traceability relative to exhaustive feature-by-feature human coding, while the human-produced transcription preserves expert musical evidence. 